\documentclass[12pt,a4paper]{article}
\pdfoutput=1
\usepackage{jheppub}
\usepackage{dsfont}
\usepackage{amsmath,amssymb,amscd,amsfonts}
\DeclareMathOperator{\Tr}{Tr}
\DeclareMathOperator{\Sinh}{Sinh}
\DeclareMathOperator{\Cosh}{Cosh}

\DeclareMathOperator{\Coth}{Coth}

\usepackage[toc,page]{appendix}
\usepackage{epsfig}
\usepackage{epstopdf}
\usepackage{latexsym}
\usepackage{graphicx}
\usepackage{subfigure}
\usepackage{booktabs}
\usepackage{bbm}
\usepackage{color}
\usepackage{physics}
\usepackage{tikz}
\usetikzlibrary{matrix}
\usetikzlibrary{decorations.markings,calc,shapes}
\usetikzlibrary{positioning}

\makeatletter
\def\@fpheader{\relax}
\makeatother

\subheader{\begin{flushright}
UTTG-16-19\\
\end{flushright}}

\title{A Killing Vector Treatment of Multiboundary Wormholes}
\author{Elena Caceres$^{a}$, Arnab Kundu$^{b}$, Ayan K.~Patra$^{b}$, Sanjit Shashi$^{a}$}
\affiliation{$^a$Theory Group, Department of Physics, University of Texas, Austin, TX 78712, USA.}
\affiliation{$^b$Theory Division, Saha Institute of Nuclear Physics, HBNI, 1/AF Bidhannagar, Kolkata 700064, India.}

\emailAdd{elenac[at]utexas.edu, sshashi[at]utexas.edu} 
\emailAdd{arnab.kundu [at] saha.ac.in, ayan.patra[at]saha.ac.in}

\abstract{
The two-sided BTZ black hole has been instrumental in elucidating several aspects of AdS/CFT. Similarly, multiboundary wormholes provide a useful and rich arena in which probing questions of quantum gravity can be posed and explored. In this work, we find the explicit forms of the Killing vectors needed to construct three-boundary wormholes, with and without rotation, as quotients of AdS$_3$. We ensure that our method captures the full moduli space of such wormholes and elaborate on the generalization of our procedure to more exotic multiboundary spaces, including higher genus.
}

\begin{document}
	
	\maketitle
	\flushbottom

	%%%%%%%%%%%%%%%%%
	\section{Introduction}
	%%%%%%%%%%%%%%%%%

	In the AdS/CFT correspondence, the bulk theory is generically a quantum gravity theory on an AdS background; in the most famous examples, it is a bona fide string theory. However, in the semiclassical limit, the partition function of the bulk theory becomes a path integral over a special class of geometries. On-shell, these geometries are obtained as the asymptotically-AdS saddles of a classical gravitational theory, such as the Einstein-Hilbert action. Nevertheless, the exact controllable computation of the complete path integral is subtle, with various issues to address. First, it is non-trivial to find {\it all} saddles in general. Secondly, one has to introduce an appropriate measure for the path integral. Furthermore, one needs to prescribe how to define the path integral meaningfully.
	
	While we will not discuss all of these subtleties here, note that, to some extent, each issue can be well-treated in a toy example with gravity in three spacetime dimensions. In particular, since such a system lacks any true dynamical bulk degrees of freedom, it is rather {\it simple} to construct all of the saddles of the action by acting with appropriate diffeomorphisms at will. We will primarily deal with this aspect\footnote{Note that similar analyses have appeared in various other works earlier, {\it e.g.}~\cite{Skenderis:2009ju, Balasubramanian:2014hda}. Building on these approaches, we will describe a complementary method for the same.} in this article. The prescription to define the path integral is, however, not limited to three spacetime dimensions; one computes the Euclidean path integral, assuming that the Lorentzian geometry has a smooth Euclidean section, and then uses analytic continuation to arrive at the Lorentzian answer. The latter part certainly assumes that such an analytic continuation exists.
%Finally, the measure for the path integral can be provided using the first observation; one has to define a {\it natural} measure on this space of inequivalent saddles as a measure on the space of the special diffeomorphisms that can take one classical saddle to another. 
	
	As, in three spacetime dimensions, dynamical bulk degrees of freedom are lacking, only global topological data and boundary dynamics classify a given classical saddle. For smooth asymptotically AdS$_3$ spaces, this implies that all such geometries are locally equivalent, differing only in their global features. There are two ingredients in the global data assuming no singularities (such as conical defects): the genus and the number of boundaries. Any asymptotically AdS$_3$ geometry has a conformal Minkowski boundary, on which the CFT is defined. Thus, in the context of AdS/CFT, the classical geometries, characterized by the global topological data, relate to the study of the $n$-fold tensor product of the CFT, where $n$ is the number of boundaries. The special case of $n=2$ corresponds to the two-sided BTZ black hole geometry, which is dual to the \textit{thermofield double (TFD)} state of the CFT\cite{Maldacena:2001kr}. The TFD state is a bipartite entangled state in the CFT, where the entangled degrees of freedom reside on the two conformal boundaries of the eternal BTZ geometry. The $n$-fold tensor product state subsequently represents a multipartite CFT state; its entanglement properties were explored in \cite{Balasubramanian:2014hda}.   
	
	In recent years, dynamical aspects have seen a surge of activities, especially in the context of holography. The standard lore of addressing dynamical questions in a QFT (or in a CFT) is the in-in formalism, also known as the Schwinger-Keldysh framework. This formalism  requires one to prepare states with the Euclidean path integral, then to use them as the ``in" state or as the ``out" state. One then glues these Euclidean states with a Lorentzian time evolution. In brief, the formalism requires evolution on a particular complex time contour. The holographic descriptions of the Euclidean section and of the Lorentzian section of the QFT are, respectively, described by a corresponding Euclidean and Lorentzian geometry. In \cite{Skenderis:2009ju}, this prescription has been described in detail, in the context of holography.\footnote{See also \cite{Skenderis:2008dh, Skenderis:2008dg} for a more detailed discussion on the real-time dictionary in holography. There are several applications and discussions on this topic, which we will not attempt to enlist here.} It is necessary, in this framework, to have multiboundary geometries in both Euclidean and in Lorentzian signatures and, subsequently, to glue them across a surface of zero extrinsic curvature.    
	
	In recent work on two-dimensional quantum gravity models, multiboundary geometries with wormholes have made an explicit appearance.	The structure in three dimensions is definitely richer and more complex, but the basic ingredients may have a similar qualitative role to play in the bigger picture.  Additionally, multiboundary wormholes provide an arena to study a wide variety of phenomena, from multipartite entanglement\cite{Bao:2015bfa}, to complexity in spaces with $n$ asymptotic regions and arbitrary internal topology\cite{Fu:2018kcp}, to traversability using a double trace deformation\cite{Gao:2016bin,Maldacena:2017axo,Fu:2018oaq, Caceres:2018ehr}.

	With all the motivations above, we present an alternative, global method of constructing multiboundary geometries with arbitrary genus in three-dimensional gravity with a negative cosmological constant. Our construction differs from the ones already described in the literature \cite{Brill:1995jv, Aminneborg:1997pz, Brill:1998pr} in several ways. For example, the construction in \cite{Skenderis:2009ju} is not global, and, without the clear foliation into hyperbolic planes provided by Poincar\'e coordinates for AdS$_3$, hard to extend to other spaces. On the other hand, \cite{Balasubramanian:2014hda} describes a somewhat different approach by identifying geodesics, but without providing explicit Killing vectors. In this work, we obtain the Killing vector needed for such a global construction of a three-boundary wormhole. Furthermore, we provide a simple algorithm by which more exotic multiboundary wormholes may be obtained.

	Our results will allow for the study of multiboundary spaces in warped AdS$_3$ and can be useful for further investigations of holographic complexity of formation\cite{Fu:2018kcp}. Also, for Euclidean spaces, there is a theorem stating that any geodesically complete space of constant negative curvature is a quotient of Euclidean AdS by a discrete subgroup of O$(3,1)$. However, to the best of our knowledge, there is no proof of such a theorem for Lorentzian signature. Prior to this work, the only known example was the two-sided case: the BTZ black hole. Our results constitute the first explicit construction of a three-boundary Lorentzian space as an AdS quotient.
	
	This article is organized as follows. Section \ref{geomAdSH} is a brief review of the geometric structure of AdS$_3$, Riemann surfaces, and Killing vectors of $\mathbb{H}$. In Section \ref{BTZquotients}, we review the quotient of the two-sided BTZ setting up notation and intuition that will be useful in later sections. Sections \ref{3bdryConst} and \ref{genConst} are the main part of this paper. In Section \ref{3bdryConst}, we present the explicit Killing vectors for the three-boundary, zero genus, case and compute the horizon lengths. In Section \ref{genConst}, we elaborate on generalizations of our methods, thereby including more boundaries, higher genus, and rotation. We conclude with directions for future research in Section \ref{conclusion}.

%%%%%%%%%%%%%%%%%%%%%%%%%
\section{Geometric Structure of AdS$_3$ \& ${\mathbb H}$}\label{geomAdSH}
%%%%%%%%%%%%%%%%%%%%%%%%%

%%%%%%%%%%%%%%%%%%%%%%%%%
\subsection{Isometries of AdS$_3$ \& Quotients}\label{quotients}
%%%%%%%%%%%%%%%%%%%%%%%%%

Let us begin with an introductory review of AdS$_3$ and its isometries. Recall that AdS$_3$ can be defined as a surface in the flat $2+2$-dimensional spacetime,
\begin{equation}
ds^2 = -d\bar{v}^2 - d\bar{u}^2 + d\bar{x}^2 + d\bar{y}^2 \ ,
\end{equation}
In particular, it is the hyperboloid surface,
\begin{equation}
-\bar{v}^2 - \bar{u}^2 + \bar{x}^2 + \bar{y}^2 = -\ell^2 \ ,
\end{equation}
where $\ell$ is the AdS curvature scale.

Taking a four-vector $\bar{x}^a = (\bar{v},\bar{u},\bar{x},\bar{y})$, AdS$_3$ has six linearly independent Killing vectors which generate rotations and boosts in the $2+2$ spacetime. These are,
\begin{equation}
J_{ab} = \bar{x}_b \frac{\partial}{\partial\bar{x}^a} - \bar{x}_a \frac{\partial}{\partial\bar{x}^b} \ .
\end{equation}
Together, these Killing vectors form an so$(2,2)$ algebra,
\begin{equation}
[J_{ab},J_{cd}] = \eta_{ac}J_{bd} - \eta_{ad}J_{bc} - \eta_{bc}J_{ad} + \eta_{bd}J_{ac} \ . \label{comm}
\end{equation}

From this point on, we will be using the Poincar\'e metric for AdS$_3$,
\begin{equation}
\frac{ds^2}{\ell^2} = \frac{-dt^2 + dx^2 + dy^2}{y^2} \ . \label{Pmet}
\end{equation}
In this metric, the Killing vectors are as follows:
\begin{align}
J_{01} &= \left(\frac{\ell^2 + t^2 + x^2 + y^2}{2\ell}\right)\partial_t + \frac{tx}{\ell}\partial_x + \frac{ty}{\ell}\partial_y \ , \label{k1} \\
J_{02} &= \left(\frac{-\ell^2 + t^2 + x^2 + y^2}{2\ell}\right)\partial_t + \frac{tx}{\ell}\partial_x + \frac{ty}{\ell}\partial_y \ , \\
J_{03} &= -x\partial_t - t\partial_x \ , \\
J_{12} &= -t\partial_t - x\partial_x - y\partial_y \ , \\
J_{13} &= \left(\frac{\ell^2-t^2-x^2+y^2}{2\ell}\right)\partial_x - \frac{tx}{\ell}\partial_t - \frac{xy}{\ell}\partial_y \ , \\
J_{23} &= \left(\frac{-\ell^2-t^2-x^2+y^2}{2\ell}\right)\partial_x - \frac{tx}{\ell}\partial_t - \frac{xy}{\ell}\partial_y \ . \label{k6}
\end{align}

In the literature, multiboundary wormholes are typically thought of as quotients of AdS$_3$ by a discrete set of isometries in the group SO$(2,2)$. We can write isometries in the identity component of SO$(2,2)$, which is isomorphic to SL$(2,\mathbb{R}) \times {\rm SL}(2,\mathbb{R})$, by exponentiating the Killing vectors. For example, dilitation comes from $J_{12}$,
\begin{equation}
e^{-2\pi\kappa J_{12}} \cdot (t,x,y) = e^{2\pi\kappa}(t,x,y) \ . \label{dilAct}
\end{equation}

When we talk about quotienting ``by an isometry" or ``by a Killing vector," we are actually identifying all points in the orbit of some group action. In particular, for a Killing vector $\xi$, we refer to the subgroup $\{e^{t\xi}\}_{t \in \mathbb{R}}$ as the \textit{one-parameter subgroup} of SO$(2,2)$ generated by $\xi$. Then, fixing a particular $t_0 \in \mathbb{R}$, we call $\{e^{t_0 \xi}\}$ the \textit{identification subgroup}. For example, once again using $\xi = -\kappa J_{12}$ and (\ref{dilAct}), the set of isometries of the form $\{e^{-2\pi n\kappa J_{12}}\}_{n \in \mathbb{Z}}$\footnote{As is done in \cite{Banados:1992gq}, we take $t_0 = 2\pi$.} is the relevant identification subgroup, and we make the identification,
\begin{equation}
(t,x,y) \sim e^{-2\pi\kappa J_{12}} \cdot (t,x,y) \implies (t,x,y) \sim e^{2\pi\kappa} (t,x,y)  \ . \label{ibtz}
\end{equation}

If everything remains well-behaved throughout quotienting, it should yield a well-defined spacetime which is locally AdS$_3$ --- the Riemann tensor and its associated quantities, all of which are defined locally, will be unchanged. However, there are possible pathologies that may arise when performing the procedure. For example, the isometry could have fixed points, so, in the quotient space, the curvature at such points would not be well-defined. In the multiboundary wormholes we consider, this is not be a problem. As we will show in Section \ref{orientationRev}, if we restrict our attention to the $t = 0$ slice, the fixed points will lie outside of the \textit{fundamental domain} of the quotient space. Thus, the curvature at every point in the multiboundary wormhole is well-defined.

Another possible pathology is the manifestation of \textit{closed timelike curves} (CTCs). If we consider a Killing vector $\xi$ which is timelike ($\xi \cdot \xi < 0$) at a particular point, then the image of that point under the corresponding finite isometry is timelike-separated. Thus, we would have a timelike curve connecting the two points, and this curve would become a CTC upon quotienting. Furthermore, observe that the Killing horizon $\xi \cdot \xi = 0$ separates the region with CTCs from the region where $\xi \cdot \xi > 0$, so we identify the Killing horizon as a causal singularity.

This issue of CTCs is overcome simply by excising the causally problematic regions.\footnote{See \cite{Banados:1992gq} for a detailed discussion on excision.} Excision eliminates the causal pathologies, but it also results in \textit{geodesic incompleteness}, since, in general, there should exist geodesics which go from the $\xi \cdot \xi > 0$ region into the $\xi \cdot \xi < 0$ region. However, although these geodesics are cut-off, they also all reach the Killing horizon of $\xi$, \textit{i.e.}, they all hit the causal singularity. Thus, the fact that these geodesics are cut-off prior to performing the quotienting does not matter.

At this point, we require a method to classify different quotient spaces of AdS$_3$ corresponding to multiboundary wormholes. For this, we use the technology of Riemann surfaces, which provides a simple way of visualizing the different possible topologies of these quotient spaces.\\

%%%%%%%%%%%%%%%%%%%%%%%%%
\subsection{Riemann Surfaces: A Brief Discussion}\label{riemann}
%%%%%%%%%%%%%%%%%%%%%%%%%

As the $t = 0$ slice of AdS$_3$ is simply the hyperbolic plane $\mathbb{H}$ (modeled by the upper half-plane), it follows that the corresponding $t = 0$ slice of a multiboundary wormhole is a quotient of $\mathbb{H}$ by a \textit{Fuchsian group}, a term for any discrete subgroup of the isometry group PSL$(2,\mathbb{R})$ of $\mathbb{H}$. In particular, all \textit{connected, hyperbolic Riemann surfaces} of genus $g$ and with $n$ boundaries, which we denote by $(n,g)$, can be written as such quotients of $\mathbb{H}$, meaning that any such Riemann surface can be used to describe the $t = 0$ slice of some multiboundary wormhole.

For these particular wormholes, the corresponding Riemann surface will retain some information about both the topology and the geometry of the total spacetime manifold. Specifically, a wormhole with $n$ boundaries and genus $g$ will be a Riemann surface $(n,g)$ at $t = 0$. The moduli space of such Riemann surfaces is known to be the \textit{Teichm\"uller space} $T(n,g)$, which, when parameterized by \textit{Fenchel-Nielsen coordinates}, is seen to be isomorphic to,
\begin{equation}
T(n,g) \cong \begin{cases}
\mathbb{R}_+,&\text{if}\ n = 2,g = 0\\
\mathbb{R}_+^{3g-3+2n} \times \mathbb{R}^{3g-3+n},&\text{if otherwise}
\end{cases} \ . \label{teichmuller}
\end{equation}
In other words, $(2,0)$ has one (positive) geometrical parameter. In all other cases, however, $(n,g)$ has $6g-6+3n$ geometrical parameters. While $3g - 3 + 2n$ of these parameters are lengths of minimal (within homotopy classes), non-intersecting, periodic geodesics, the other $3g - 3 + n$ parameters are all ``twist" angles.\footnote{See \cite{Skenderis:2009ju} for a more detailed discussion of these parameters.}

At $t = 0$, the black hole horizons become minimal periodic geodesics on the Riemann surface.\footnote{\cite{Skenderis:2009ju} explicitly shows that the metric outside of these minimal periodic geodesics is precisely that of the static BTZ.} With regards to the geometrical parameters, this means that $n$ of the positive parameters correspond to the lengths (or masses) of the horizons seen from the exterior regions of the wormhole, while the remaining parameters describe the internal geometry behind these horizons. If we restrict our attention to static wormholes---that is, those with zero angular momentum---then the Riemann surface has all of the parameters of the wormhole geometry, which is made apparent from the metric found by \cite{Skenderis:2009ju}.

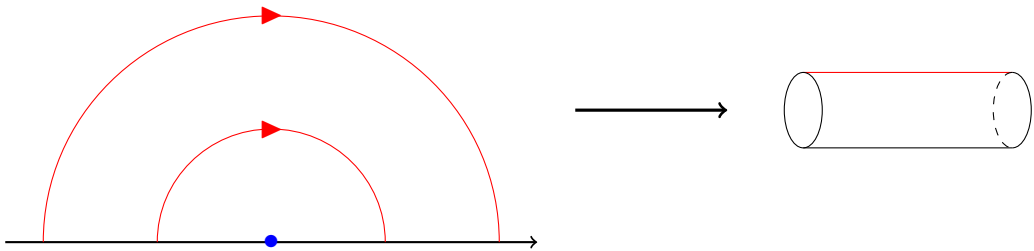
\begin{figure}
	\centering
	\begin{tikzpicture}
	\draw[->,thick] (-3.5,0) to (3.5,0);
	\draw[-,color=red] (-1.5,0) arc (180:0:1.5);
	\draw[-,color=red] (3,0) arc (0:180:3);
	
	\node[color=red,rotate=-90] at (0,1.5) {$\blacktriangle$};
	\node[color=red,rotate=-90] at (0,3) {$\blacktriangle$};
	
	\draw[->,very thick] (4,1.75) to (6,1.75);
	
	\draw[-,color=red] (7,2.25) to (9.75,2.25);
	\draw[-] (7,1.25) to (9.75,1.25);
	\draw[-] (7,1.75) ellipse (0.25 and 0.5);
	\draw[-] (9.75,2.25) arc (90:-90:0.25 and 0.5);
	\draw[-,dashed] (9.75,2.25) arc (90:270:0.25 and 0.5);
	
	\node[color=blue] at (0,0) {$\bullet$};
	\end{tikzpicture}
	\caption{The Riemann surface obtained by quotienting the upper half-plane by dilatation (\ref{ibtz}). The fundamental domain is bounded by the red semicircles, while the fixed point of (\ref{ibtz}) is the center of the semicircles. This is the $t = 0$ slice of the two-sided BTZ.}
	\label{figs:rs2side}
\end{figure}

The constructions of \cite{Skenderis:2009ju} and \cite{Balasubramanian:2014hda} make heavy use of Riemann surfaces. Especially, \cite{Skenderis:2009ju} focuses first on constructing the $(n,g)$ Riemann surface, then \textit{lifting} the isometries that they use from $\mathbb{H}$ to AdS$_3$ in order to construct the $(n,g)$-wormhole metric beyond the $t = 0$ slice. Physically, this lifting step can be viewed as time-evolving the Riemann surface without altering the topological data. Thus, the relevant Riemann surface provides a visualization of the corresponding wormhole. Indeed, this procedure of constructing the Riemann surface as the $t = 0$ slice first allows us to see that the identification of equation (\ref{ibtz}) yields the two-sided BTZ (Figure \ref{figs:rs2side}).

Nevertheless, \cite{Skenderis:2009ju} has a very ``piecewise" feature to the construction. For instance, in the $(3,0)$ (\textit{three-boundary}) case, for which the corresponding Riemann surface is a pair of pants, we know from \cite{Balasubramanian:2014hda} that the appropriate procedure would be to quotient by two isometries, using two Killing vectors. Meanwhile, in \cite{Skenderis:2009ju}, they cover the pair of pants with three overlapping cylinders, construct those from subspaces of $\mathbb{H}$, and glue the resulting three metrics together in order to reconstruct the pair of pants. In short, \cite{Skenderis:2009ju} does not describe a global quotienting procedure of AdS$_3$.

This patchy, piecewise method is perfectly fine if we only wish to consider AdS$_3$, since it relies heavily on the presence of a hyperbolic $t = 0$ slice. However, in other cases, such as that of warped AdS$_3$,\footnote{See \cite{Anninos:2008fx,Jugeau:2010nq} for a more detailed discussion of warped AdS$_3$.} since there is not a clear idea of a sufficient hyperbolic slice of the spacetime, we cannot repeat this procedure. A more global construction is needed should one attempt to construct such a multiboundary warped AdS$_3$ geometry. Specifically, given the Killing vectors used to construct an $(n,g)$-wormhole in AdS$_3$, it should be possible to extrapolate the Killing vectors needed for the analogous warped AdS$_3$ wormhole. So, for now, our goal is to explicitly write the two global Killing vectors which would yield the three-boundary AdS$_3$ solution.

There are some details regarding the overall procedure that require attention. First, as mentioned in Section \ref{quotients}, we need to ensure that there are no fixed points inside of the fundamental domain of the Riemann surface. For instance, in the case of dilatation, the fixed point is the origin; as seen in Figure \ref{figs:rs2side}, this point is not contained within the fundamental domain of the Riemann surface. In Section \ref{orientationRev} will confirm that this holds for any Riemann surface to ensure smoothness.

Additionally, there is some ambiguity in the lifting. As we will see in Section \ref{KVsinH}, some of the Killing vectors vanish on the $t = 0$ slice. Thus, the lifted Killing vector could include a linear combination of such terms. However, these terms would introduce new parameters for the wormhole geometry; as discussed in \cite{Banados:1992gq}, the rotating two-sided BTZ includes such a term. So, if we only consider static wormholes, we exclude such terms when lifting in order for the parameter counting to make sense.

%%%%%%%%%%%%%%%%%%%%%%%%%
\subsection{Killing Vectors of ${\mathbb H}$}\label{KVsinH}
%%%%%%%%%%%%%%%%%%%%%%%%%

Let us go back to the Killing vectors above. Prior discussion in this section makes it clear that we will need to have the Killing vectors of $\mathbb{H}$. Setting $t = 0$ in equations (\ref{k1})--(\ref{k6}) yields these vectors; denoting them as $J_{ab}^{(0)}$, we write:
\begin{align}
J_{01}^{(0)} &= J_{20}^{(0)} = J_{03}^{(0)} = 0 \ , \label{k1to3H}\\
J_{12}^{(0)} &= -x\partial_x - y\partial_y \ , \label{k4H}\\
J_{13}^{(0)} &= \left(\frac{\ell^2-x^2+y^2}{2\ell}\right)\partial_x - \frac{xy}{\ell}\partial_y \ , \label{k5H}\\
J_{23}^{(0)} &= \left(\frac{-\ell^2-x^2+y^2}{2\ell}\right)\partial_x - \frac{xy}{\ell}\partial_y \ . \label{k6H}
\end{align}
As expected of a maximally-symmetric, two-dimensional space, we have three linearly independent Killing vectors, corresponding to three independent isometries.

Now, our goal is to write another more convenient basis of Killing vectors of $\mathbb{H}$ in terms of $J_{12}^{(0)}$, $J_{13}^{(0)}$, and $J_{23}^{(0)}$. Afterwards, we will use the new basis Killing vectors in order to explicitly write isometries which yield the three-boundary geometry described in \cite{Skenderis:2009ju, Balasubramanian:2014hda}. Subsequently, we will verify that quotienting by these isometries yields three independent geometrical parameters by checking the lengths of the non-intersecting, minimal, periodic geodesics, which are related to their respective horizon masses by,
\begin{equation}
L = 2\pi\ell\sqrt{M}
\end{equation}

First, note that we can combine (\ref{k5H}) and (\ref{k6H}) to take two of the three independent Killing vectors to be,
\begin{equation}
J_{13}^{(0)} + J_{23}^{(0)} = \left(\frac{-x^2 + y^2}{\ell}\right)\partial_x - \frac{2xy}{\ell} \partial_y \ ,\ \ J_{13}^{(0)} - J_{23}^{(0)} = \ell \partial_x \ . 
\end{equation}
At this stage, it helps to describe the upper half-plane in terms of complex coordinates $(z,\bar{z})$ instead of $(x,y)$,
\begin{equation}
z = x + iy\ ,\ \ \bar{z} = x - iy \ . \label{complexCoords}
\end{equation}
Thus, we get that:
\begin{align}
J_{12}^{(0)} &= -z\partial_z - \bar{z}\partial_{\bar{z}}  \ , \\
J_{13}^{(0)} + J_{23}^{(0)} &=  \frac{1}{\ell}\left(-z^2 \partial_z - \bar{z}^2 \partial_{\bar{z}}\right) \ , \\
J_{13}^{(0)} - J_{23}^{(0)} &= \ell\left(\partial_{z} + \partial_{\bar{z}}\right) \ . 
\end{align}

Now, we define the following basis Killing vectors of $\mathbb{H}$, also noting which Killing vectors they lift to in AdS$_3$ in the static case
\begin{eqnarray}
J_{T} = \ell\left(\partial_z + \partial_{\bar{z}}\right) & \longrightarrow& \tilde{J}_{T} = J_{13} - J_{23} \ , \\
J_{D} = z\partial_z + \bar{z}\partial_{\bar{z}} & \longrightarrow& \tilde{J}_{D} = -J_{12} \ , \\
J_{S} = \dfrac{1}{\ell} \left(z^2\partial_z + \bar{z}^2\partial_{\bar{z}}\right) & \longrightarrow& \tilde{J}_{S} = -J_{13} - J_{23} \ .
\end{eqnarray}
Here the tilde stands for the Killing vector lifted to AdS$_3$. Using (\ref{comm}), we confirm that $\tilde{J}_T$, $\tilde{J}_D$, and $\tilde{J}_S$ form an sl$(2,\mathbb{R})$ subalgebra of so$(2,2)$.
\begin{align}
[\tilde{J}_D,\tilde{J}_T] &= -\tilde{J}_T \ , \label{subalg1}\\
[\tilde{J}_D,\tilde{J}_S] &= \tilde{J}_S \ , \label{subalg2}\\
[\tilde{J}_S,\tilde{J}_T] &= -2\tilde{J}_D \ .\label{subalg3}
\end{align}

Each Killing vector should generate a particular linear fractional transformation in PSL$(2,\mathbb{R})$ which can be reduced to the identity. Indeed, by exponentiating each of them with some dimensionless parameter $\kappa$ and acting on a point $z$ in $\mathbb{H}$, we have:
\begin{align}
e^{\kappa J_T} \cdot z
&= \sum_{n=0}^{\infty} \frac{(\kappa\ell)^n}{n!} (\partial_z + \partial_{\bar{z}})^n \cdot z
= z + \kappa\ell \ , \\
e^{\kappa J_D} \cdot z
&= \sum_{n=0}^{\infty} \frac{\kappa^n}{n!} (z\partial_z + \bar{z}\partial_{\bar{z}})^n \cdot z
= \sum_{n=0}^{\infty} \frac{\kappa^n}{n!} z
= e^{\kappa}z \ , \\
e^{\kappa J_S} \cdot z
&= \sum_{n=0}^{\infty} \frac{\kappa^n}{n!\ell^n} (z^2 \partial_z + \bar{z}^2 \partial_{\bar{z}})^n \cdot z
= \sum_{n=0}^{\infty} \frac{\kappa^n}{n!\ell^n} (n! z^{n+1})
= \frac{z}{-\kappa z/\ell + 1} \ . 
\end{align}

where each $\kappa \in \mathbb{R}$. Now, we rewrite these three, independent isometries in terms of the $(x,y)$ coordinates ({\it i.e.}, the real and imaginary components of the $z$ coordinate), which more clearly allows us to identify them as \textit{translation}, \textit{dilatation}, and a \textit{special conformal transformation (SCT)}, respectively.
\begin{align}
e^{\kappa J_T} \cdot (x,y) &= (x+\kappa\ell,y)\ , \label{trans} \\
e^{\kappa J_D} \cdot (x,y) &= e^{\kappa}(x,y)\ , \label{dila} \\
e^{\kappa J_S} \cdot (x,y) &= \frac{\ell^2}{(\ell-\kappa x)^2 + (\kappa y)^2} \left(x - \frac{\kappa}{\ell}(x^2 + y^2),y\right)\ . \label{sct}
\end{align}

We have not yet discussed inversions, which are defined as,
\begin{equation}
z \to -\frac{\alpha}{z} \ .
\end{equation}
For this, $[\alpha] = \ell^2$. We should be able to write this isometry as a composition of translations, dilatations, and SCTs. Indeed, define the following operator,
\begin{equation}
\mathcal{I}_a = e^{aJ_T}e^{J_S/a}e^{aJ_T} \ , \label{inversion}
\end{equation}
where $a$ is a dimensionless parameter. We apply $\mathcal{I}_a$ to $z$.
\begin{align*}
\mathcal{I}_a \cdot z = e^{aJ_T} e^{J_S/a} e^{aJ_T} \cdot z
&= e^{aJ_T} e^{J_S/a} \cdot (z + a\ell)\\
&= e^{aJ_T} \cdot \left[\frac{z + a\ell}{-(z + a\ell)/(a\ell) + 1}\right]\\
&= e^{aJ_T} \cdot \left[-\frac{(a\ell)^2}{z} - a\ell\right]\\
&= -\frac{(a\ell)^2}{z} \ .
\end{align*}

Let us now discuss how these isometries act on the geodesics of $\mathbb{H}$. First, it is straightforward to find the geodesics in $\mathbb{H}$ by noting that such curves will extremize the following length-functional,
\begin{eqnarray}
I = \int \frac{dy}{y}\sqrt{1 + \left( \frac{dx}{dy}\right)^2}  \ ,
\end{eqnarray}
Using the Euler-Lagrange equation, the corresponding equation of motion is,
\begin{eqnarray}
\frac{1}{y} \frac{dx}{dy} \frac{1}{\sqrt{1 + (dx/dy)^2}} = \alpha \ , \label{eom}
\end{eqnarray}
where $\alpha$ is an integral of motion. There are two distinct classes of solutions to (\ref{eom}),
\begin{align}
\alpha = 0 \ , &\quad x = {\rm constant} \ , \label{geodesic1}\\
\alpha \not = 0 \ , &\quad (x-c)^2 + y^2 = \alpha^{-2} \ ,\label{geodesic2}
\end{align}
\textit{i.e.}, straight lines and semicircles.

Now, we compute how the isometries act on the semicircular geodesics. For such a geodesic centered at $(c,0)$ and radius $R$, translation yields,
\begin{equation}
e^{aJ_T/\ell} \cdot \left(x_1,\sqrt{R^2 - (x_1 - c)^2}\right) = \left(x_2,\sqrt{R^2 - (x_2 - a - c)^2}\right),\ \ \ x_2 = x_1 + a \ . \label{tcirc}
\end{equation}
So, translation parameterized by a length $a$ will shift semicircles to the right by $a$.

A similar process can performed with dilatation, which yields,
\begin{equation}
e^{aJ_D}\left(x_1,\sqrt{R^2 - (x_1 - c)^2}\right) = \left(x_2,\sqrt{(e^a R)^2 - (x_2 - e^{a}c)^2}\right),\ \ \ x_2 = e^{a}x_1 \ . \label{dcirc}
\end{equation}
As expected, dilatation will change both the radius and the center of a semicircle by a scale factor $e^a$. Semicircles at the origin thus only change in size.

Finally, we specifically consider what inversion does to semicircles centered at the origin. To do so, note that inversion in $(x,y)$ coordinates is,
\begin{equation}
\mathcal{I}_a \cdot (x,y) = \frac{(a\ell)^2}{x^2+y^2}(-x,y) \ . \label{icirc1}
\end{equation}
Applying this to a semicircle of radius $R$ centered at the origin, we thus have,
\begin{equation}
\mathcal{I}_a \cdot \left(x_1,\sqrt{R^2 - x_1^2}\right) = \left(x_2,\sqrt{\frac{(a\ell)^4}{R^2} - x_2^2}\right),\ \ \ x_2 = -\frac{(a\ell)^2}{R^2} x_1 \ . \label{icirc2}
\end{equation}
Note that, in the case $a = R/\ell$, the corresponding inversion operator simply flips the orientation of the semicircle. Mathematically,
\begin{equation}
\mathcal{I}_{R/\ell} \cdot \left(x_1,\sqrt{R^2-x_1^2}\right) = \left(x_2,\sqrt{R^2 - x_2^2}\right),\ \ x_2 = -x_1 \ .
\end{equation}
At this stage, we are well-equipped to construct isometries with which we can describe the identification of two (and more) semicircles to create the multiboundary geometry.

%%%%%%%%%%%%%%%%%%%%%%%%%
\section{Global Quotients of the Two-Sided BTZ}\label{BTZquotients}
%%%%%%%%%%%%%%%%%%%%%%%%%

%%%%%%%%%%%%%%%%%%%%%%%%%
\subsection{Orientation-Reversing Isometries}\label{orientationRev}
%%%%%%%%%%%%%%%%%%%%%%%%%

\begin{figure}
	\centering
	\begin{tikzpicture}[scale=0.7]
	\draw[->,thick] (-3.5,0) to (3.5,0);
	\draw[-,color=blue] (1,0) arc (180:0:0.75/2);
	\draw[-,color=blue] (2.75,0) arc (0:180:0.75/2);
	
	\draw[-,color=red] (0.75,0) arc (0:180:0.75);
	\draw[-,color=red] (3,0) arc (0:180:3);
	
	\node[color=blue,rotate=-90] at (1+0.75/2,0.75/2) {$\blacktriangle$};
	\node[color=blue,rotate=90] at (2.75-0.75/2,0.75/2) {$\blacktriangle$};
	
	\node[color=red,rotate=-90] at (0,0.75) {$\blacktriangle$};
	\node[color=red,rotate=-90] at (0,3) {$\blacktriangle$};
	
	\draw[->,very thick,dashed,red] (0,1) to (0,2.75);
	\draw[->,very thick,dashed,blue] (1+0.75/2,0.6) to[bend left] (2.75-0.75/2,0.6);

	\draw[->,very thick] (4,1.75) to (6,1.75);

	\draw[-,color=red] (7,2.75) to (9.75,2.75);
	\draw[-] (7,0.75) to (9.75,0.75);
	\draw[-] (7,1.75) ellipse (0.25 and 1);
	\draw[-] (9.75,2.75) arc (90:-90:0.25 and 1);
	\draw[-,dashed] (9.75,2.75) arc (90:270:0.25 and 1);
	
	\draw[-,dashed,blue] (9.5,1.5) .. controls (8.5,1.7) and (8.5,1.8) .. (9.5,2);
	\draw[-,blue] (10,1.5) .. controls (9,1.7) and (9,1.8) .. (10,2);
	
	\node[rotate=-180,color=blue] at (8.75,1.75) {$\blacktriangle$};
	\node[rotate=-180,color=blue] at (9.25,1.75) {$\blacktriangle$};

	\draw[->,very thick] (11,1.75) to (13,1.75);

	\draw[-,color=red] (7+7,2.25) to[bend right] (7+9,3);
	\draw[-,color=blue] (7+9.7,2.28) .. controls (7+9,3.49/2) .. (7+9.7,1.21);
	
	\draw[-] (7+7,1.75) ellipse (0.25 and 0.5);
	\draw[-] (7+7,1.25) to[bend left] (7+9,0.5);
	
	\draw[-,rotate around={45:(7+9,2.5)}] (7+9.35,2.35) ellipse (0.25 and 0.5);
	\draw[-,rotate around={135:(7+9,0.5)}] (7+9,0) ellipse (0.25 and 0.5);

	\draw[->,thick] (-3.5,0-4) to (3.5,0-4);
	\draw[-,color=red!30!yellow] (2.75,0-4) arc (0:180:0.75);
	\draw[-,color=red!30!yellow] (-2.75,0-4) arc (180:0:0.75);
	
	\draw[-,color=red] (0.75,0-4) arc (0:180:0.75);
	\draw[-,color=red] (3,0-4) arc (0:180:3);
	
	\node[color=red!30!yellow,rotate=90] at (2,0.75-4) {$\blacktriangle$};
	\node[color=red!30!yellow,rotate=-90] at (-2,0.75-4) {$\blacktriangle$};
	
	\node[color=red,rotate=-90] at (0.,0.75-4) {$\blacktriangle$};
	\node[color=red,rotate=-90] at (0,3-4) {$\blacktriangle$};
	
	\draw[->,very thick,dashed,red] (0,0.7-4) to (0,2.75-4);
	\draw[->,very thick,dashed,red!30!yellow] (-2,1-4) to[bend left] (2,1-4);

	\draw[->,very thick] (4,1.75-4) to (6,1.75-4);

	\draw[-,color=red] (7,2.75-4) to (9.75,2.75-4);
	\draw[-] (7,0.75-4) to (9.75,0.75-4);
	\draw[-] (7,1.75-4) ellipse (0.25 and 1);
	\draw[-] (9.75,2.75-4) arc (90:-90:0.25 and 1);
	\draw[-,dashed] (9.75,2.75-4) arc (90:270:0.25 and 1);
	
	\draw[-,red!30!yellow] (7.25,1.5-4) .. controls (8.25,1.7-4) and (8.25,1.8-4) .. (7.25,2-4);
	\draw[-,red!30!yellow] (10,1.5-4) .. controls (9,1.7-4) and (9,1.8-4) .. (10,2-4);
	
	\node[rotate=-180,red!30!yellow] at (8,1.75-4) {$\blacktriangle$};
	\node[rotate=-180,red!30!yellow] at (9.25,1.75-4) {$\blacktriangle$};

	\draw[->,very thick] (11,1.75-4) to (13,1.75-4);

	\draw[-] (7+7,1.75-4) ellipse (0.25 and 0.5);
	\draw[-,color=red] (7+7,2.25-4) .. controls (7+7.75,2-4) .. (7+8.5,2.5-4) arc(130:-130:1) .. controls (7+7.75,1.5-4) .. (7+7,1.25-4);
	
	\draw[-] (7+8.9,1.75-4) arc (200:340:0.25);
	\draw[-] (7+8.855,1.68-4) arc (160:20:0.3);
	
	\draw[-,color=red!30!yellow] (7+7.225,1.875-4) .. controls (7+7.225/2+8.9/2,2-4) .. (7+8.9,1.75-4);
	\draw[-,color=red!30!yellow,dashed] (7+8.9,1.75-4) .. controls (7+7.225/2+8.9/2,1.5-4) .. (7+6.775,1.625-4);
	\end{tikzpicture}
	\caption{The three-boundary and one-boundary, one-genus Riemann surfaces as quotients of the two-boundary Riemann surface. The three-boundary surface is obtained by ``\textit{pinching}" one of the boundaries into two, while the one-boundary, one-genus surface is obtained by ``\textit{folding}" one of the boundaries onto the other.}
	\label{figs:pinchfold}
\end{figure}
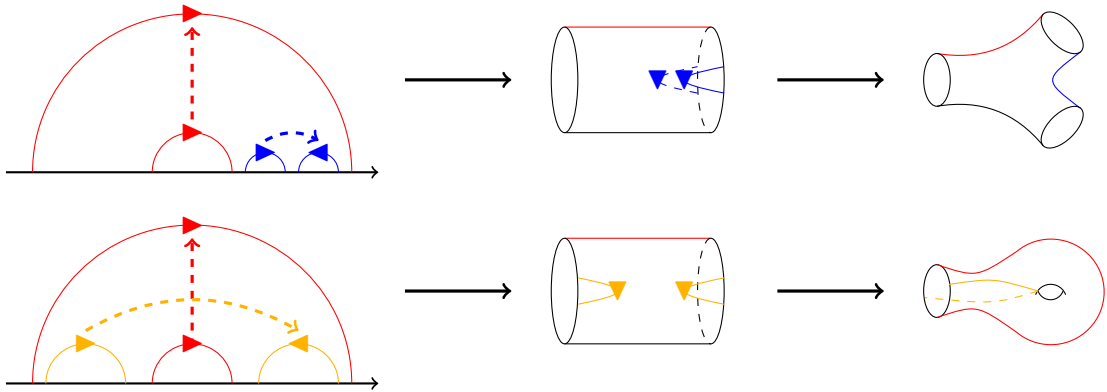

Before we proceed, let us briefly discuss the necessary identification in a purely pictorial sense. Figure \ref{figs:rs2side} already depicts the identification used to obtain the two-sided BTZ, which is a cylinder at $t = 0$. However, we can further quotient this cylinder in order to obtain more exotic Riemann surfaces, as shown in Figure \ref{figs:pinchfold}.\footnote{\cite{Balasubramanian:2014hda} shows pictorially that the three-boundary case is a quotient of the two-boundary case; we will ultimately extend this idea to other geometries.}

This works topologically, and the combination of pinching and folding can even be used to construct Riemann surfaces with any number of boundaries and any genus. We will discuss this further in Section \ref{genConst}.

However, Figure \ref{figs:pinchfold} is not a very explicit picture. All that we can see from it is that the isometry by which we quotient the two-sided BTZ in order to pinch a boundary has the same sort of action as that which we use to fold two boundaries---both isometries reverse the orientation of some semicircle, then translate it elsewhere. Thus, in order to better understand this procedure, we will first write this type of isometry explicitly.

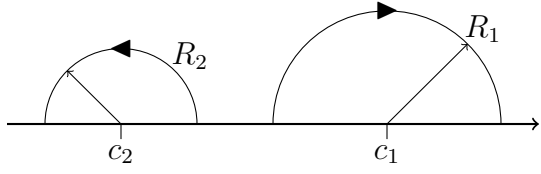
\begin{figure}
	\centering
	\begin{tikzpicture}
	\draw[->,thick] (-3.5,0) to (3.5,0);
	
	\draw[-] (3,0) arc (0:180:1.5);
	\draw[-] (-3,0) arc (180:0:1);
	
	\node[rotate=-90] at (1.5,1.5) {$\blacktriangle$};
	\node[rotate=90] at (-2,1) {$\blacktriangle$};
	
	\draw (1.5,0) to (1.5,-0.2);
	\draw (-2,0) to (-2,-0.2);
	
	\draw[->] (-2,0) to (-2-1.414/2,1.414/2);
	\draw[->] (1.5,0) to (1.5+1.414*1.5/2,1.414*1.5/2);
	
	\node at (1.5+1.414*1.5/2+0.2,1.414*1.5/2+0.2) {$R_1$};
	\node at (-2+1.414/2+0.2,1.414/2+0.2) {$R_2$};
	
	\node at (1.5,-0.4) {$c_1$};
	\node at (-2,-0.4) {$c_2$};
	\end{tikzpicture}
	\caption{The action of the general orientation-reversing isometry on two arbitrary semicircular geodesics in $\mathbb{H}$. Also defined are the centers and radii of the semicircles.}
	\label{figs:orientationRev}
\end{figure}

Consider the picture in Figure \ref{figs:orientationRev}. We will refer to the semicircle on the right as $C_1$ and the semicircle on the left as $C_2$. Our goal will be to transform $C_1$ into $C_2$ as shown.

We know that orientation of a semicircle of radius $R$ centered at the origin can be reversed by applying inversion, which is defined in (\ref{inversion}), with $a = R/\ell$. It is reasonable to think that we need to apply $\mathcal{I}_{R_{1}/\ell}$ at some point. As such, we will work with the following composition of isometries, also using (\ref{trans})-(\ref{sct}),
\begin{equation}
\mathcal{O} = e^{c_2 J_T/\ell} e^{\nu J_D} \mathcal{I}_{R_{1}/\ell} e^{-c_1 J_T/\ell} \ . \label{KillingO}
\end{equation}
Let us define the $\nu$ parameter to be,
\begin{equation}
\nu = \log\left(\frac{R_2}{R_1}\right) \implies e^{\nu} = \frac{R_2}{R_1} \ .
\end{equation}

So, (\ref{orientationRev}) takes $C_1$ to the origin, flips its orientation, dilates the newly-flipped semicircle, and translates the result to match $C_2$. We can confirm this sequence works by applying each isometry, step-by-step, to an arbitrary point on $C_1$, using (\ref{tcirc}), (\ref{dcirc}), and (\ref{icirc2}).
\begin{align}
e^{-c_1 J_T} \cdot \left(x_1,\sqrt{R_1^2 - (x_1 - c_1)^2}\right)
&= \left(x_2,\sqrt{R_1^2 - x_2^2}\right) \ , \nonumber\\
\mathcal{I}_{R_1/\ell} \cdot \left(x_2,\sqrt{R_1^2-x_2^2}\right)
&=  \left(x_3,\sqrt{R_1^2 - x_3^2}\right) \ , \nonumber\\
e^{\nu J_D} \cdot \left(x_3,\sqrt{R_1^2-x_3^2}\right)
&=  \left(x_4,\sqrt{R_2^2-x_4^2}\right) \ , \nonumber\\
e^{c_2 J_T} \cdot \left(x_4,\sqrt{R_2^2-x_4^2}\right)
&= \left(x_5,\sqrt{R_2^2-(x_5-c_2)^2}\right) \ , \nonumber\\
\implies \mathcal{O} \cdot \left(x_1,\sqrt{R_1^2-(x_1-c_1)^2}\right)
&= \left(x_5,\sqrt{R_2^2-(x_5-c_2)^2}\right) \ . \label{circFlip}
\end{align}
To ensure that orientation is truly flipped, we write $x_5$ in terms of $x_1$.
\begin{align}
&x_5 = x_4 + c_2 = \frac{R_2}{R_1}x_3 + c_2 = -\frac{R_2}{R_1}x_2 + c_2 = -\frac{R_2}{R_1}(x_1 - c_1) + c_2 \ , \nonumber\\
&\implies x_5 = \frac{R_2}{R_1}(c_1-x_1) + c_2 \label{inv}
\end{align}
We can use (\ref{inv}) to show that the left (right) half of $C_1$ indeed maps onto the right (left) half of $C_2$.
\begin{align*}
c_1 \leq x_1 \leq c_1 + R_1
&\implies c_2 \geq x_5 \geq c_2 - R_2 \ , \\
c_1 \geq x_1 \geq c_1 - R_1
&\implies c_2 \leq x_5 \leq c_2 + R_2 \ .
\end{align*}
Thus, the transformation in (\ref{KillingO}) is an \textit{orientation-reversing isometry} that could be used in both pinching and folding.

However, as discussed in Section \ref{geomAdSH}, we need to check that the fixed points of ($\ref{KillingO}$) are contained within the semicircles. If there are fixed points, then the quotient space would not be smooth; we would expect some sort of orbifold singularity on which the curvature is not well-defined.

We find the fixed points in terms of the complex coordinate, defined in (\ref{complexCoords}). Applying (\ref{KillingO}) to a fixed point $z$ yields,
\begin{equation}
z = \mathcal{O} \cdot z = -\frac{e^\nu R_1^2}{z - c_1} + c_2 = -\frac{R_1 R_2}{z-c_1} + c_2 \ . \label{fixedpoint1}
\end{equation}
If we solve for $z$, then we get that there are two fixed points,
\begin{equation}
z_{\pm} = \frac{1}{2}\left[c_1 + c_2 \pm \sqrt{(c_1 - c_2)^2 - 4R_1 R_2}\right]\ .\label{fixedpointvals}
\end{equation}
As in Figure \ref{figs:orientationRev}, we impose the inequality,
\begin{equation}
|c_1 - c_2| > R_1 + R_2\ . \label{orientationRevIneq}
\end{equation}
From this inequality, we have that the terms in the square root of (\ref{fixedpointvals}) are strictly positive, implying that the fixed points are real.

Because the fixed points are real, they lie somewhere along the $x$-axis. In fact, utilizing (\ref{orientationRevIneq}), we can show that the fixed points lie within the semicircles; for Figure 3, in which $c_1 > c_2$, $z_+$ lies in $C_1$, while $z_-$ lies in $C_2$. Thus, in the fundamental domain of a quotient space which involves the identification of such semicircles, the fixed points always lie outside, meaning that the curvature is well-defined everywhere.

So, to summarize, the isometry (\ref{KillingO}) is an appropriate, orientation-reversing isometry, and quotienting by it provides a manifold with well-defined curvature. Taking a quotient by this sequence of transformations is an explicit presentation of both the pinching and folding procedures, so we can use (\ref{KillingO}) to understand cases beyond the two-sided BTZ. In particular, we will use a specific version of this isometry to form the three-boundary wormhole.

%%%%%%%%%%%%%%%%%%%%%
\subsection{Picturing the Three-Boundary Wormhole}\label{picturing3sided}
%%%%%%%%%%%%%%%%%%%%%

%
\begin{figure}
	\[
	\begin{tikzpicture}
	\draw[->,very thin] (-5.5,0) to (5.5,0);
	\draw[->,very thin] (0,0) to (0,5.5);
	
	\draw[-,dashed,very thick,color=red] (0,1.25) to (0,5);
	\draw[-,dashed,very thick,color=blue] (41/12,0.372678) arc (41.81:138.19:0.559017);
	\draw[-,dashed,very thick,color=red!65!blue] (39/20,2/5) arc (36.87:118.07:2/3);
	\draw[-,dashed,very thick,color=red!65!blue] (4*75/68,4*10/17) arc (162.02:176.99:7.62209);
	
	\draw[-,color=red] (5,0) arc (0:180:5);
	\draw[-,color=red] (1.25,0) arc (0:180:1.25);
	
	\draw[-,color=blue] (1.75,0) arc (180:0:0.5);
	\draw[-,color=blue] (3.25,0) arc (180:0:0.5);
	
	\node[color=red,rotate=-90] at (0,1.25) {$\blacktriangle$};
	\node[color=red,rotate=-90] at (0,5) {$\blacktriangle$};
	
	\node[color=blue,rotate=-90] at (2.25,0.5) {$\blacktriangle$};
	\node[color=blue,rotate=90] at (3.75,0.5) {$\blacktriangle$};
	
	\draw (0,0) to (0,-0.2);
	\draw (2.25,0) to (2.25,-0.2);
	\draw (3.75,0) to (3.75,-0.2);
	
	\node at (0,-0.4) {$0$};
	\node at (2.25,-0.4) {$c_1$};
	\node at (3.75,-0.4) {$c_2$};
	
	\draw[->] (0,0) to (1.25/1.414,1.25/1.414);
	\node at (1.25/1.414-0.3,1.25/1.414+0.5) {$R_0$};
	
	\draw[->] (0,0) to (-5/1.414,5/1.414);
	\node at (-5/1.414+0.7,5/1.414-0.2) {$\lambda R_0$};
	
	\draw[->] (2.25,0) to (2.25-0.5/1.414,0.5/1.414);
	\node at (2.25,0.5/1.414+0.5) {$R$};
	
	\draw[->] (3.75,0) to (3.75+0.5/1.414,0.5/1.414);
	\node at (3.75,0.5/1.414+0.5) {$R$};
	
	\node at (0.3,3.125) {$L_1$};
	\node at (3,0.8) {$L_2$};
	\node at (1.5,1) {$L_{3}^L$};
	\node at (3.9,2) {$L_3^R$};
	\end{tikzpicture}
	\]
	\caption{The fundamental domain of the three-boundary Riemann surface. The color-coded dashed lines $L_{1,2,3}$ are the minimal periodic geodesics, whose lengths are the three \textit{physical parameters} of the system. The variables $\lambda$, $R_0$, $R$, $c_1$, and $c_2$ represent parameters for the picture.}
	\label{figs:3bdrypic}
\end{figure}
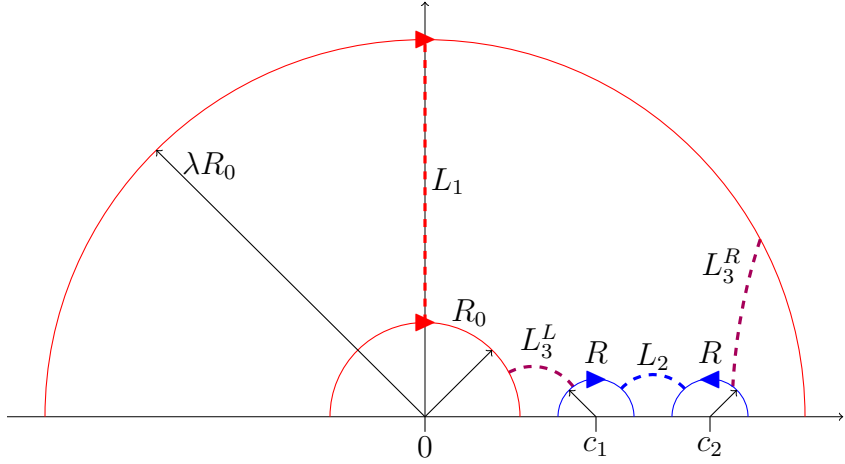

Figure \ref{figs:3bdrypic} is a more explicit version of the first picture in Figure \ref{figs:pinchfold}. First, notice that we have drawn several dashed geodesics (denoted by $L_{1,2,3}$); after performing the appropriate identifications, these color-coded geodesics become closed (\textit{i.e.}, \textit{periodic}) curves. Furthermore, if we impose that these geodesics be \textit{minimal} within their homotopy classes ({\it i.e.}, within the class of periodic geodesics which can be continuously transformed into each other by sliding them along the Riemann surface), then they can be identified as the black hole horizons.\footnote{\cite{Skenderis:2009ju} shows that, in the wormhole geometry, the metric outside of the causal development of these minimal periodic geodesics is simply the BTZ metric in \cite{Banados:1992gq}.}

Note that we have introduced specific parameters. We will hereby refer to these as the \textit{identification parameters}; they will appear in the coefficients of the Killing vectors that exponentiates to the appropriate isometry. Furthermore, we have specifically chosen both of the blue semicircles to have the same radii. We could have taken them to have different radii, but we will show that the three physical parameters are independent even in this picture. Thus, Figure \ref{figs:3bdrypic} captures the full moduli space of three-boundary static wormholes.\\

Before we proceed, note that we will often alternate between using $c_1$ and $c_2$ and using their sum and difference, which are also independent of one another.
\begin{align}
s &= c_2 + c_1 \ ,\label{sDef}\\
d &= c_2 - c_1 \ .\label{dDef}
\end{align}

%%%%%%%%%%%%%%%%%%%%%%%%%%%%%%%%
\section{Static Three-Boundary Construction} \label{3bdryConst}
%%%%%%%%%%%%%%%%%%%%%%%%%%%%%%%%

%%%%%%%%%%%%%%%%%%%%%%%%%%%%%%%%
\subsection{Killing Vectors for Three-Boundaries} \label{KVfor3bdry}
%%%%%%%%%%%%%%%%%%%%%%%%%%%%%%%%

In light of the discussion of Section \ref{BTZquotients}, the three-boundary wormhole is obtained by quotienting AdS$_3$ by dilatation, then by the orientation-reversing isometry depicted in Figure \ref{figs:3bdrypic},
\begin{equation}
\tilde{\mathcal{O}} = e^{c_2 \tilde{J}_T/\ell} e^{R \tilde{J}_T/\ell} e^{\ell \tilde{J}_S/R} e^{R \tilde{J}_T/\ell} e^{-c_1 \tilde{J}_T/\ell}\ .\label{3bdrySecondID}
\end{equation}
However, this is a finite transformation. We thus discuss how to derive the infinitesimal version, which is a Killing vector of the initial AdS$_3$ geometry.

We intend to write (\ref{3bdrySecondID}) as a single exponential. Because the Killing vectors $\{\tilde{J}_T,\tilde{J}_D,\tilde{J}_S\}$ form a subalgebra (described in (\ref{subalg1})-(\ref{subalg3})), the Baker-Campbell-Hausdorff formula implies, taking $a,b,c \in \mathbb{R}$,
\begin{equation}
\tilde{\mathcal{O}} = e^{a\tilde{J}_T + b\tilde{J}_D + c\tilde{J}_S} = e^{\xi_{3B}}\ .\label{singleExp}
\end{equation}

Considering the $t = 0$ slice of AdS$_3$, we find the action of this operator on a point $z \in \mathbb{H}$ as follows. We temporarily set $\ell = 1$; $\ell$ will be unnecessary in the final result. First, consider the infinitesimal shift obtained by acting on $z$ by the generator $\xi_{3B}$,
\begin{equation}
\delta z = \xi_{3B} \cdot z = \left(a + bz + cz^2\right) = -\bar{a}\left[(z - \bar{b})^2 - \bar{c}^2\right] \ .\label{infTrans}
\end{equation}
where we define the coefficients $\bar{a}$, $\bar{b}$, and $\bar{c}$ by:
\begin{align}
a &= \bar{a}(\bar{c}^2 - \bar{b}^2) \ ,\label{aCoeff}\\
b &= 2\bar{a}\bar{b} \ ,\label{bCoeff}\\
c &= -\bar{a} \ .\label{cCoeff}
\end{align}

Note that $\bar{a}$ and $\bar{b}$ are real, but $\bar{c}$ can be real or imaginary.\footnote{The sign of $\bar{c}^2$ is directly related to the \textit{type} of the Killing vector (discussed in \cite{Banados:1992gq}). This can be seen from the Casimir invariants of $\xi_{3B}$; they are $I_1 = -8\bar{a}^2\bar{c}^2$ and $I_2 = 0$.} Defining $z'$ as the image of $z$ under the finite transformation, we rewrite (\ref{infTrans}) and integrate to obtain,
\begin{equation}
\delta\left[\log\left(\frac{z-\bar{b}-\bar{c}}{z-\bar{b}+\bar{c}}\right)\right] = \log\left[\left(\frac{z'-\bar{b}-\bar{c}}{z'-\bar{b}+\bar{c}}\right)\left(\frac{z-\bar{b}+\bar{c}}{z-\bar{b}-\bar{c}}\right)\right] = -\bar{a}\bar{c}
\end{equation}
We are then able to solve for $z'$ to find that the finite transformation takes the form,
\begin{equation}
z' = \bar{b} + \bar{c} \frac{(z-\bar{b})\cosh(\bar{a}\bar{c}) + \bar{c}\sinh(\bar{a}\bar{c})}{(z-\bar{b})\sinh(\bar{a}\bar{c}) + \bar{c}\cosh(\bar{a}\bar{c})} = \bar{b} + \frac{(z-\bar{b})\cosh(\bar{a}\bar{c}) + \bar{c}\sinh(\bar{a}\bar{c})}{(z-\bar{b})\sinh(\bar{a}\bar{c})/\bar{c} + \cosh(\bar{a}\bar{c})}\ .\label{finTrans}
\end{equation}

Thus, the right-hand side of (\ref{finTrans}) is a linear fractional transformation with real coefficients (even if $\bar{c}$ is imaginary) and determinant $1$, making it an element of the isometry group PSL$(2,\mathbb{R})$. However, for the static three-boundary construction, which only needs two independent parameters, we may impose the constraint,\footnote{We will briefly discuss (\ref{finTrans}) without enforcing (\ref{constraintFinTrans}) in Appendix \ref{appA}.}
\begin{equation}
\bar{c}^2 = 1 \implies \bar{c} = \pm 1\ .\label{constraintFinTrans}
\end{equation}
Imposing (\ref{constraintFinTrans}) reduces (\ref{finTrans}) to,
\begin{equation}
z' = \bar{b} + \frac{(z-\bar{b})\cosh(\bar{a}) + \sinh(\bar{a})}{(z-\bar{b})\sinh(\bar{a}) + \cosh(\bar{a})}\ .\label{finTrans2}
\end{equation}

Now, we also demand that (\ref{finTrans2}) match the isometry depicted in Figure \ref{figs:3bdrypic}. Symbolically, in $(x,y)$ coordinates, if we take an arbitrary point on the semicircle $y = \sqrt{R^2 - (x-c_1)^2}$, this action is,
\begin{equation}
(x',y') = (c_1 + c_2 - x,y)\ .\label{actionCirc}
\end{equation}

Using (\ref{sDef}) and (\ref{dDef}), (\ref{finTrans2}) and (\ref{actionCirc}) precisely match if we have,
\begin{align}
s &= 2\bar{b} \ ,\label{sVal}\\
d &= 2\coth(\bar{a}) \ ,\label{dVal}
\end{align}
with an additional constraint on $d$ and $R$,
\begin{equation}
\frac{d^2}{4} - R^2 = 1\ .\label{dRConst}
\end{equation}

Now, using (\ref{aCoeff})-(\ref{cCoeff}), we can explicitly write the Killing vector coefficients which appear in (\ref{singleExp}) as follows:
\begin{align}
a &= \Coth^{-1}\left(\frac{d}{2}\right)\left(1 - \frac{s^2}{4}\right) \ ,\label{aSolved}\\
b &= s\Coth^{-1}\left(\frac{d}{2}\right)\ ,\label{bSolved}\\
c &= -\Coth^{-1}\left(\frac{d}{2}\right)\ .\label{cSolved}
\end{align}

Before presenting the Killing vector $\xi_{3B}$ in full, we use (\ref{dRConst}) to write its coefficients in terms of ratios $d/R$ and $s/R$. The $\ell$ thus becomes unnecessary, as these parameters are dimensionless. Specifically, solve for $R$ in terms of $d/R$ to write,
\begin{align}
d &= \frac{d}{R}\frac{2}{\sqrt{\left(\frac{d}{R}\right)^2 - 4}} \ ,\label{dAndRatio}\\
s &= \frac{s}{R}\frac{2}{\sqrt{\left(\frac{d}{R}\right)^2 - 4}} \ .\label{sAndRatio}
\end{align}
Hence, the Killing vector in (\ref{singleExp}) is,
\begin{align}
\xi_{3B}
= &\Coth^{-1}\left[\frac{d}{R}\frac{1}{\sqrt{\left(\frac{d}{R}\right)^2 - 4}}\right]\left[1 - \left(\frac{s}{R}\right)^2 \frac{1}{\left(\frac{d}{R}\right)^2 - 4}\right] \tilde{J}_T\nonumber\\
& + \left[\frac{s}{R}\frac{2}{\sqrt{(\frac{d}{R})^2 - 4}}\right]\Coth^{-1}\left[\frac{d}{R}\frac{1}{\sqrt{\left(\frac{d}{R}\right)^2 - 4}}\right]\tilde{J}_D\nonumber\\
& - \Coth^{-1}\left[\frac{d}{R}\frac{1}{\sqrt{\left(\frac{d}{R}\right)^2 - 4}}\right]\tilde{J}_S\ .\label{KV3B}
\end{align}

Observe that (\ref{KV3B}) has two independent, dimensionless parameters: $d/R$ and $s/R$. This is as expected according to (\ref{teichmuller}). While the two-sided BTZ only has one geometrical parameter, the three-boundary wormhole should have three, with the two new parameters being in $\xi_{3B}$.

Additionally, in the language of \cite{Banados:1992gq}, this vector is type I$_b$. This can be checked by computing the Casimir invariants of $\xi_{3B}$; one is $0$ and the other is negative.

%%%%%%%%%%%%%%%%%%%%%%%%%%%%%%%%
\subsection{Horizon Length in Terms of Identification Parameters}\label{horiLen}
%%%%%%%%%%%%%%%%%%%%%%%%%%%%%%%%

For the sake of completeness, we will verify that the identification of Figure \ref{figs:3bdrypic} captures the full moduli space of three-boundary wormholes at $t = 0$. In other words, our goal is to determine that the three horizons, computed as minimal periodic geodesics, are indeed independent of one another.

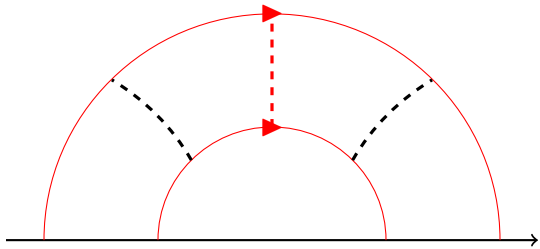
\begin{figure}
	\centering
	\begin{tikzpicture}
	\draw[->,thick] (-3.5,0) to (3.5,0);
	\draw[-,color=red] (-1.5,0) arc (180:0:1.5);
	\draw[-,color=red] (3,0) arc (0:180:3);
	
	\draw[-,dashed,color=red,very thick] (0,1.5) to (0,3);
	\draw[-,dashed,color=black,very thick] (1.5/1.414,1.5/1.414) arc (150:121:2.98);
	\draw[-,dashed,color=black,very thick] (-1.5/1.414,1.5/1.414) arc (180-150:180-121:2.98);

	\node[color=red,rotate=-90] at (0,1.5) {$\blacktriangle$};
	\node[color=red,rotate=-90] at (0,3) {$\blacktriangle$};	
	\end{tikzpicture}
	\caption{The Riemann surface obtained by quotienting the upper half-plane by dilatation (\ref{ibtz}), with three of the resulting periodic geodesics drawn. By symmetry, we can see that the vertical geodesic is extremal.}
	\label{figs:dilatationPeriodic}
\end{figure}

We start with $L_1$, which, in Figure \ref{figs:3bdrypic}, is simply a vertical line from $(0,R_0)$ to $(0,\lambda R_0)$. To see why, note that identifying the geodesics related by dilatation yields a cylindrical geometry, and the geodesics which become periodic are either circular arcs or the line $x = 0$, as shown in Figure \ref{figs:dilatationPeriodic}. Furthermore, the symmetry of this picture indicates that the periodic $x = 0$ geodesic is shorter than the other periodic geodesics.

The length of this line can be found using (\ref{Pmet}), as follows.
\begin{equation}
L_1 = \ell \int_{R_0}^{\lambda R_0} \frac{dy}{y} = \ell\log\left(\frac{\lambda R_0}{R_0}\right) = \ell\log\lambda \ . \label{loglamb}
\end{equation}
This minimal periodic geodesic corresponds to the horizon of the two-sided BTZ. Additionally, it corresponds to one of the horizons of the three-boundary wormhole, since it is homotopic to one of the conformal boundaries in Figure \ref{figs:3bdrypic}.

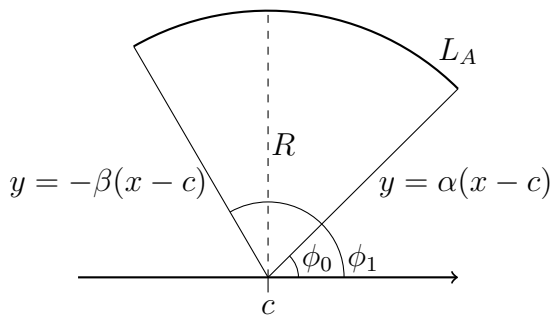
\begin{figure}
\centering
\begin{tikzpicture}
\draw[->,thick] (-2.5,0) to (2.5,0);
\draw[-] (0,0) to (2.5,2.5);
\draw[-] (0,0) to (-1.767766,3.06186);
\node at (2.6,1.25) {$y = \alpha (x-c)$};
\node at (-2.1,1.25) {$y = -\beta (x-c)$};
\draw[-] (0,0) to (0,-0.2);
\node at (0,-0.4) {$c$};
\draw[-,dashed] (0,0) to (0,5/1.414);
\node at (0.2,2.5/1.414) {$R$};

\draw[-,thick] (2.5,2.5) arc (45:120:5/1.414);

\draw[-] (0.4,0) arc (0:45:0.4);
\node at (0.65,0.25) {$\phi_0$};
\draw[-] (1,0) arc (0:120:1);
\node at (1.25,0.25) {$\phi_1$};

\node at (2.5,3) {$L_A$};
\end{tikzpicture}
\caption{A circular arc centered at a point $c$ and of radius $R$ in $\mathbb{H}$. Lines of slope $1/\alpha$ and $-1/\beta$ bound the arc and make angles of $\phi_0$ and $\phi_1$, respectively, with the $x$-axis. $L_A$ is the length of the arc.}
\label{figs:circArc}
\end{figure}

For the other horizons, we need the length of a circular arc centered at a point on the $x$-axis ({\it i.e.}, the length of portions of the circular geodesics in $\mathbb{H}$). Consider Figure \ref{figs:circArc}.

In this picture, we have assumed $\alpha,\beta > 0$ for the sake of simplicity, but the following argument works as long as $\alpha$ is the slope of the right bounding line while $-\beta$ is the slope of the left bounding line. We can parameterize the arc itself in terms of the angle $\phi$,
\begin{equation}
x(\phi) = R\cos\phi + c,\quad y(\phi) = R\sin\phi\quad (\phi_0 \leq \phi \leq \phi_1) \ . 
\end{equation}
Plugging this into the metric in (\ref{Pmet}) (again with $t = 0$) yields,
\begin{equation}
L_A
= \ell \int_{\phi_0}^{\phi_1} d\phi\ \sqrt{\frac{R^2 \sin^2\phi + R^2 \cos^2\phi}{R^2 \sin^2\phi}}
= \ell \log\left[\tan\left(\frac{\phi_1}{2}\right)\cot\left(\frac{\phi_0}{2}\right)\right] \ . \label{arcLength}
\end{equation}

Observe that $\tan\phi_0 = \alpha$ and $\tan(\pi-\phi_1) = -\tan\phi_1 = \beta$. Additionally, we can use trigonometric identities and the fact that $0 \leq \phi_0 \leq \pi/2 \leq \phi_1 \leq \pi$ to write,
\begin{equation}
\tan\left(\frac{\phi_1}{2}\right)\cot\left(\frac{\phi_0}{2}\right) = \left(\sqrt{1+\frac{1}{\beta^2}} + \frac{1}{\beta}\right) \left(\sqrt{1 + \frac{1}{\alpha^2}} + \frac{1}{\alpha}\right) \ . \label{trigAngs}
\end{equation}
Combining (\ref{trigAngs}) with (\ref{arcLength}) yields,
\begin{equation}
\frac{L_A}{\ell} = \Sinh^{-1}\left(\frac{1}{\alpha}\right) + \Sinh^{-1}\left(\frac{1}{\beta}\right) \ . \label{arcLength2}
\end{equation}

Equation (\ref{arcLength2}) will be used quite a bit in calculating $L_2$, $L_3^L$, and $L_3^R$, as shown in Figure \ref{figs:3bdrypic}. In particular, observe that $L_A$ is independent of $R$; only the slopes of the lines determine $L_A$. Thus, in order to minimize a circular arc centered at a particular point, we need to take $\alpha$ and $\beta$ to simultaneously be as large as possible.

We now calculate $L_2$, for which we consider Figure \ref{figs:calcL2}. First, note that, for any periodic curve which connects the two blue semicircles, the endpoints must be identified when quotienting by the isometry discussed above. Specifically, we denote the endpoints of $L_2$ as $\left(x_L^{(2)},y_L^{(2)}\right)$ and $\left(x_R^{(2)},y_R^{(2)}\right)$. Furthermore, we define $c_0^{(2)}$ to be the midpoint of the two semicircles.
\begin{equation}
c_0^{(2)} = \frac{c_1 + c_2}{2} \ .\label{midpoint}
\end{equation}
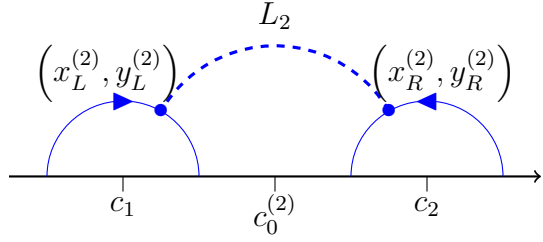
\begin{figure}
\centering
\begin{tikzpicture}
\draw[->,thick] (-3.5,0) to (3.5,0);

\draw[-,color=blue] (-1,0) arc (0:180:1);
\draw[-,color=blue] (3,0) arc (0:180:1);
\node[rotate=-90,color=blue] at (-2,1) {$\blacktriangle$};
\node[rotate=90,color=blue] at (2,1) {$\blacktriangle$};

\node[color=blue] at (3/2,1.73205/2) {$\bullet$};
\node[color=blue] at (-3/2,1.73205/2) {$\bullet$};

\draw[-,dashed,very thick,color=blue] (3/2,1.73205/2) arc (30:150:1.73205);
\node at (0,2.13205) {$L_2$};

\node at (3/2+0.7,1.73205/2+0.5) {$\left(x_R^{(2)},y_R^{(2)}\right)$};
\node at (-3/2-0.7,1.73205/2+0.5) {$\left(x_L^{(2)},y_L^{(2)}\right)$};

\draw[-] (-2,0) to (-2,-0.2);
\draw[-] (2,0) to (2,-0.2);
\draw[-] (0,0) to (0,-0.2);

\node at (-2,-0.4) {$c_1$};
\node at (2,-0.4) {$c_2$};
\node at (0,-0.5) {$c_0^{(2)}$};
\end{tikzpicture}
\caption{The semicircles identified by the orientation-reversing isometry. $L_2$ is the minimal periodic geodesic obtained through this identification, and, by symmetry, it must be centered at the midpoint of $c_1$ and $c_2$.}
\label{figs:calcL2}
\end{figure}

To impose periodicity, the left endpoint of $L_2$ must map to the right endpoint under the orientation-reversing isometry. So, using (\ref{inv}) with (\ref{circFlip}) and noting that $R_1 = R_2 = R$ in this case, we have that,
\begin{align}
x_R^{(2)} &= 2c_0^{(2)} + x_L^{(2)} \ ,\label{x2Act}\\
y_R^{(2)} &= y_{L}^{(2)} \ . \label{y2Act}
\end{align}

Thus, with the constraint that $L_2$ be a periodic geodesic, taking $\alpha_0 > 0$ to be the slope of the line connecting $\left(c_0^{(2)},0\right)$ with $\left(x_R^{(2)},y_R^{(2)}\right)$, we can use (\ref{arcLength2}) to write,
\begin{equation}
\frac{L_2}{\ell} = 2\Sinh^{-1}\left(\frac{1}{\alpha_0}\right) \ . \label{eqn2}
\end{equation}

Now, need to impose minimality, which we do by maximizing $\alpha_0$. Note that any line of positive slope which originates from $\left(c_0^{(2)},0\right)$ and intersects with the right semicircle has maximal slope if and only if there is precisely one intersection point. We can use this constraint to find the appropriate value of $\alpha_0$.

In particular, consider the following equation.
\begin{align}
&\sqrt{R^2 - (x - c_2)^2} = \alpha_0\left(x - c_0^{(2)}\right)\nonumber\\
&\implies x = \frac{2c_2 + \alpha_0^2(c_1 + c_2) \pm \sqrt{4R^2 + \alpha_0^2\left[4R^2 - (c_1 - c_2)^2\right]}}{2(1+\alpha_0^2)} \ . 
\end{align}
If there is only one intersection point, then the terms in the square root must sum to zero. Thus, we conclude that, as $\alpha_0 > 0$,
\begin{equation}
\alpha = \frac{2R}{\sqrt{(c_1 - c_2)^2 - 4R^2}} \implies \frac{1}{\alpha} = \sqrt{\frac{1}{4}\left(\frac{c_1 - c_2}{R}\right)^2 - 1} \ . \label{eqn1}
\end{equation}

We now combine (\ref{eqn1}) and (\ref{eqn2}), incorporating the definition of $d$ in (\ref{dDef}).
\begin{equation}
L_2 = 2\ell\Sinh^{-1}\left[\sqrt{\frac{1}{4}\left(\frac{d}{R}\right)^2-1}\right] \ . \label{L2val}
\end{equation}

$L_3$, the third minimal periodic geodesic obtained by combining $L_3^L$ and $L_3^R$ in Figure \ref{figs:3bdrypic}, is more difficult to solve for explicitly in terms of the identification parameters, as we will see. As such, we only confirm that it is independent of $L_1$ and $L_2$. By (\ref{loglamb}) and (\ref{L2val}), this amounts to checking that, for any function of $\lambda$, $R$, and $d$,
\begin{equation}
L_3 \neq f(\lambda,R,d)\ .\label{L3condition}
\end{equation}

To do so, we first calculate the lengths $L_3^L$ and $L_3^R$ with arbitrary endpoints, but still imposing periodicity. Denote the left and right endpoints of $L_3^L$ as $\left(x_L^{(3)},y_L^{(3)}\right)$ and $\left(x_R^{(3)},y_R^{(3)}\right)$, respectively, and consider the center of the arc at position $(c_{L},0)$. We can use the Pythagorean theorem to solve for $c_L$.
\begin{align}
&\left(x_L^{(3)} - c_L\right)^2 + \left(y_L^{(3)}\right)^2 = 
\left(x_R^{(3)} - c_L\right)^2 + \left(y_R^{(3)}\right)^2\nonumber\\
&\implies c_L = \frac{\left[\left(x_L^{(3)}\right)^2 + \left(y_L^{(3)}\right)^2\right] - \left[\left(x_R^{(3)}\right)^2 + \left(y_R^{(3)}\right)^2\right]}{2\left(x_L^{(3)} - y_L^{(3)}\right)} \ .
\end{align}

By construction, the left and right endpoints are on the semicircles $y = \sqrt{R_0^2 - x^2}$ and $y = \sqrt{R^2 - (x-c_1)^2}$, respectively. As such, we can rewrite $c_L$ as follows.
\begin{equation}
c_L = \frac{R_0^2 - R^2 + c_1^2 - 2c_1 x_R^{(3)}}{2\left(x_L^{(3)} - x_R^{(3)}\right)} \ . \label{cL}
\end{equation}
Now, using equation (\ref{arcLength2}), we have,
\begin{align}
\frac{L_3^L}{\ell}
&= \Sinh^{-1}\left(-\frac{x_L^{(3)} - c_L}{y_{L}^{(3)}}\right) + \Sinh^{-1}\left(\frac{x_R^{(3)} - c_L}{y_{R}^{(3)}}\right)\nonumber\\
&= -\Sinh^{-1}\left(\frac{x_L^{(3)} - c_L}{\sqrt{R_0^2 - \left(x_L^{(3)}\right)^2}}\right) + \Sinh^{-1}\left(\frac{x_R^{(3)} - c_L}{\sqrt{R^2 - \left(x_R^{(3)} - c_1\right)^2}}\right)\ . \label{eom2}
\end{align}

Next, we insist on periodicity. Thus, the endpoints of $L_3^R$ are given by,
\begin{align}
\left(x_L^{(3)},y_L^{(3)}\right) &\to \left(\lambda x_{L}^{(3)},\lambda y_L^{(3)}\right)  \ , \\
\left(x_R^{(3)},y_R^{(3)}\right) &\to \left(c_1 + c_2 - x_R^{(3)},y_R^{(3)}\right) \ .
\end{align}
Taking the center of this arc to be $(c_R,0)$, we can compute it just as for $c_L$ above.
\begin{align}
&\left(c_1 + c_2 - x_R^{(3)} - c_R\right)^2 + \left(y_R^{(3)}\right)^2 = \left(\lambda x_L^{(3)} - c_R\right)^2 + \left(\lambda y_L^{(3)}\right)^2\nonumber\\
&\implies c_R = \frac{\lambda^2 \left[\left(x_L^{(3)}\right)^2 + \left(y_L^{(3)}\right)^2\right]
	- \left[\left(c_1 + c_2 - x_R^{(3)}\right)^2 + \left(y_R^{(3)}\right)^2\right]}{2\left(\lambda x_L^{(3)} + x_R^{(3)} - c_1 - c_2\right)}\ .
\end{align}

Again, we use the fact that $\left(x_L^{(3)},y_L^{(3)}\right)$ is on $y = \sqrt{R_0^2 - x^2}$ and $\left(x_R^{(3)},y_R^{(3)}\right)$ is on $y = \sqrt{R^2 - (x-c_1)^2}$, but to rewrite $c_R$.
\begin{equation}
c_R = \frac{\lambda^2 R_0^2 - R^2 - c_2^2 - 2c_1 c_2 + 2c_2 x_R^{(3)}}{2\left(\lambda x_L^{(3)} + x_R^{(3)} - c_1 - c_2\right)} \ . 
\end{equation}

Now, we note that (\ref{arcLength2}) cannot be used blindly for $L_3^R$, because we have three possible cases: $\lambda x_L^{(3)} > c_1 + c_2 - x_R^{(3)}$, $\lambda x_L^{(3)} < c_1 + c_2 - x_R^{(3)}$, and $\lambda x_L^{(3)} = c_1 + c_2 - x_R^{(3)}$. As such, we need to be careful with signs. Let us treat each case separately.

For Case I, $\lambda x_L^{(3)} > c_1 + c_2 - x_R^{(3)}$, we get,
\begin{align}
\frac{L_3^{R,I}}{\ell}
&= \Sinh^{-1}\left(-\frac{c_1 + c_2 - x_R^{(3)} - c_R}{y_R^{(3)}}\right) + \Sinh^{-1}\left(\frac{\lambda x_L^{(3)} - c_R}{\lambda y_L^{(3)}}\right)\nonumber\\
&= \Sinh^{-1}\left(\frac{x_R^{(3)} + c_R - c_1 - c_2}{\sqrt{R^2 - \left(x_R^{(3)} - c_1\right)^2}}\right) + \Sinh^{-1}\left(\frac{\lambda x_L^{(3)} - c_R}{\lambda\sqrt{R_0^2 - \left(x_L^{(3)}\right)^2}}\right) \ . \label{cas1}
\end{align}

For Case II, $\lambda x_L^{(3)} < c_1 + c_2 - x_R^{(3)}$, we essentially switch the slopes which we plug into (\ref{arcLength2}) for Case I, which yields,
\begin{equation}
\frac{L_3^{R,II}}{\ell} = -\frac{L_3^{R,I}}{\ell} \ . \label{cas2}
\end{equation}

Finally, for Case III, $\lambda x_L^{(3)} = c_1 + c_2 - x_R^{(3)}$, we replicate the calculations used to obtain $L_1$ in (\ref{loglamb}), since the geodesic in this case would be a straight line.\footnote{We can also obtain $L_3^{R,III}$ from either $L_3^{R,I}$ or $L_3^{R,II}$ by taking the limit $\lambda x_L^{(3)} \to c_1 + c_2 - x_R^{(R)}$ from the right or the left, respectively.}
\begin{equation}
\frac{L_3^{R,III}}{\ell} = \log\left[\frac{\sqrt{\lambda^2 R_0^2 - \lambda^2 \left(x_L^{(3)}\right)^2}}{\sqrt{R^2 - \left(x_R^{(3)}-c_1\right)^2}} 
\right]  \ . \label{cas3}
\end{equation}

Combining (\ref{eom2}) with either (\ref{cas1}), (\ref{cas2}), or (\ref{cas3}), depending on the case we are analyzing, we ultimately write $L_3^L + L_3^R$ as,
\begin{equation}
L_3^L + L_3^R = \begin{cases}
L_3^L + \left|L_3^{R,I}\right|\ \ &\text{if}\ \lambda x_L^{(3)} \neq c_1 + c_2 - x_R^{(3)} \ , \\
L_3^L + L_3^{R,III}\ \ &\text{if}\ \lambda x_L^{(3)} = c_1 + c_2 - x_R^{(3)} \ . \label{l3leng}
\end{cases}
\end{equation}
This is the quantity which we must minimize over the $\left(x_L^{(3)},x_R^{(3)}\right)$ parameter space.

As mentioned before, this sum is more difficult to minimize analytically; the same geometric trick we used for $L_2$ will not work, because there are two pieces. Additionally, the minimum of $L_3$ does not necessarily coincide with the minima of $L_3^L$ or $L_3^R$. So, in checking the independence of $L_3$ from $L_1$ and $L_2$, we use numerical methods to validate the condition (\ref{L3condition}), which amounts to showing that $L_3$ depends on either $R_0$ or $s$ in (\ref{sDef}).

First, we check the dependence of $L_3$ on $R_0$. One may expect that $L_3$ is actually independent of $R_0$. Indeed, increasing $R_0$ while keeping all other parameters fixed will increase $L_3^R$ while decreasing $L_3^L$. Conversely, decreasing $R_0$ will increase $L_3^L$ while decreasing $L_3^R$. To check this numerically, fix $\lambda$, $R$, $c_1$, and $c_2$ as follows.\footnote{The parameters $c_1$, $c_2$, and $R$ are technically dimensionful, but we set $\ell = 1$ for convenience.}
\begin{equation}
\lambda = 4,\quad R = \dfrac{1}{2},\quad c_1 = \dfrac{9}{4}, \quad c_2 = \dfrac{15}{4} \ . \label{fixPara1}
\end{equation}

\begin{figure}
\centering
\includegraphics[scale=0.55]{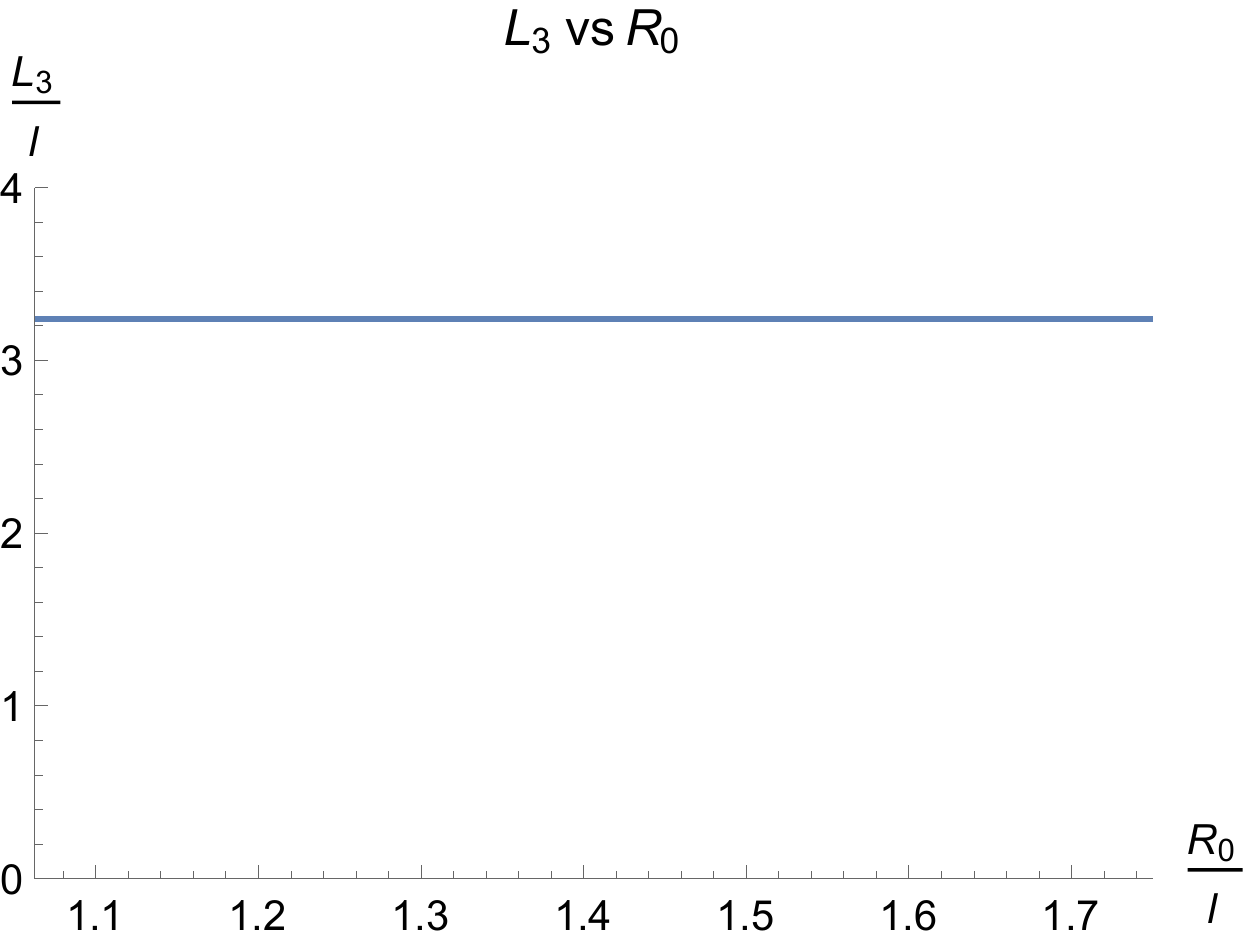}\quad \includegraphics[scale=0.55]{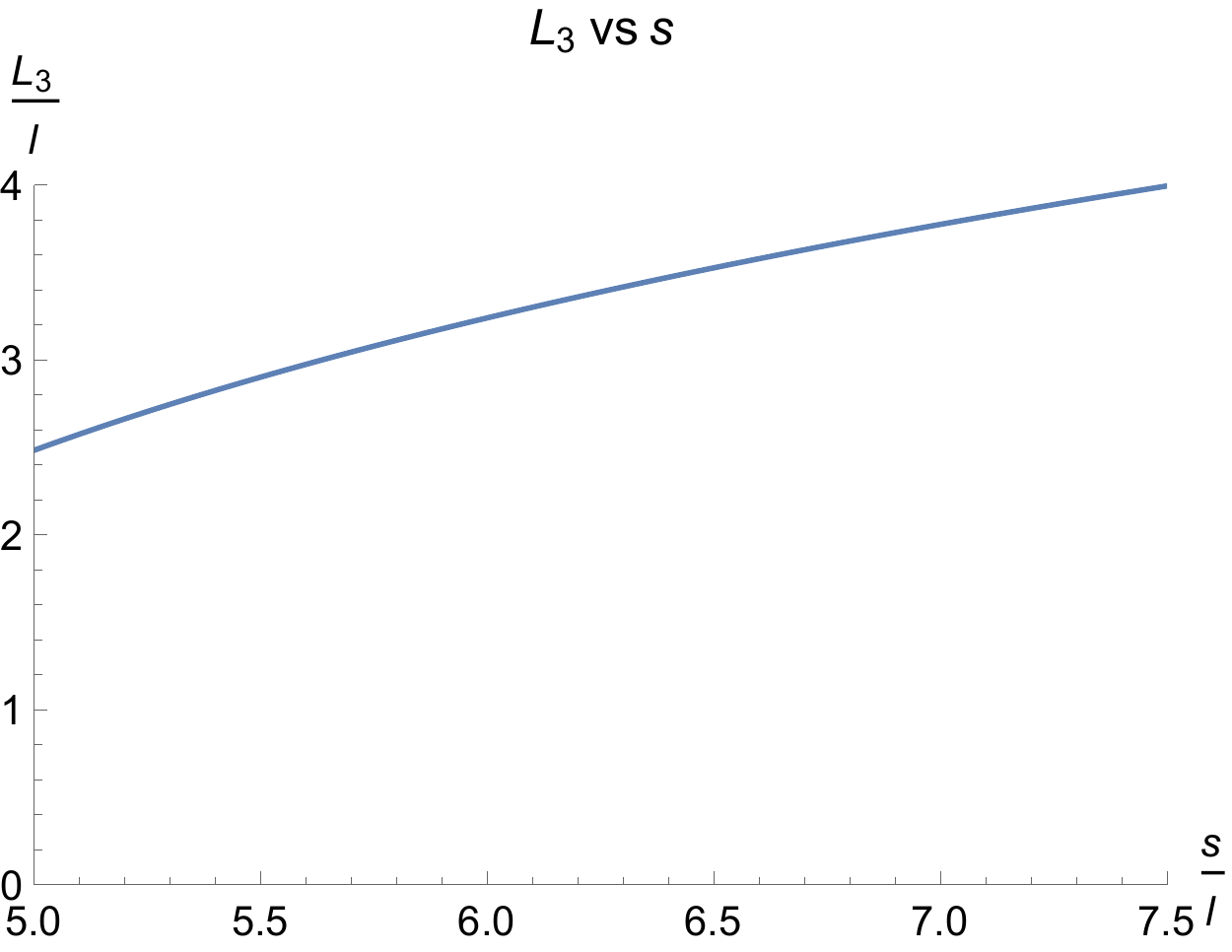}
\caption{Plots of the third minimal periodic geodesic $L_3$ as a function of $R_0$ (left) and as a function of $s$ (right), at $\ell = 1$. $L_3$ appears to depend on $s$.}
\label{figs:plotsL3}
\end{figure}

From Figure \ref{figs:3bdrypic}, we deduce that,
\begin{equation}
\frac{c_2+R}{\lambda} < R_0 < c_1-R \implies \frac{17}{16} < R_0 < \frac{28}{16} \ . \label{R0ineq}
\end{equation}
Thus, we can plug the values specified in (\ref{fixPara1}) into (\ref{l3leng}), then plot the minimum of this sum with respect to the allowed values of $R_0$. This yields the first plot in Figure \ref{figs:plotsL3}.

This plot seems to indicate that $L_3$ is independent of $R_0$. So, we now check the dependence of $L_3$ on $s$, fixing $\lambda$, $R$, $R_0$, and $d$ as follows.
\begin{equation}
\lambda = 4, \quad R = \dfrac{1}{2},\quad R_0 = \dfrac{5}{4},\quad d = \dfrac{3}{2} \ . \label{fixPara2}
\end{equation}

Just as we have bounds on $R_0$, we wish to obtain bounds on $s$, keeping the identification parameters listed in (\ref{fixPara2}) fixed. We deduce from Figure \ref{figs:3bdrypic} that,
\begin{equation}
2R_0 + 2R + d < s < 2\lambda R_0 - 2R - d \implies 5 < s < \frac{15}{2} \ . \label{sineq}
\end{equation}
Thus, we repeat the procedure used to draw the first plot in order to obtain a second one depicting the dependence of $L_3$ on $s$, also shown in Figure \ref{figs:plotsL3}. From this plot, $L_3$ can be seen to vary with $s$, which confirms the condition (\ref{L3condition}) and further verifies that the lengths $L_1$, $L_2$, and $L_3$ in Figure \ref{figs:3bdrypic} are all independent from one another.

%%%%%%%%%%%%%%%%%%%%%
\section{General Wormhole Construction}\label{genConst}
%%%%%%%%%%%%%%%%%%%%%

Let us now discuss generalizing our construction to include genus, a larger number of boundaries, and rotation.

%%%%%%%%%%%%%%%%%%%%%%%%%%%%%%%%%%
\subsection{Static (1,1) Construction} \label{static11}
%%%%%%%%%%%%%%%%%%%%%%%%%%%%%%%%%%

We will briefly review the static $(1,1)$-wormhole construction, in which we only have one boundary, but introduce genus. This particular case has been explored in the past by \cite{Aminneborg:1997pz, Aminneborg:1998si, Krasnov:2001va}. \cite{Aminneborg:1997pz} even provides a pair of Killing vectors which yields a $(1,1)$-wormhole, but they do not capture the full moduli space of solutions; their Killing vectors only have one parameter. Using our Killing vector, it is possible to attain the full count of three parameters.

\begin{figure}
	\[
	\begin{tikzpicture}
	\draw[->,very thin] (-5.5,0) to (5.5,0);
	\draw[->,very thin] (0,0) to (0,5.5);
	
	\draw[-,dashed,very thick,color=red] (0,1.25) to (0,5);

	\draw[-,dashed,very thick,color=red!30!yellow] (1.25/1.06,1.25/2.92) arc (120:45:0.6);
	\draw[-,dashed,very thick,color=red!30!yellow] (-3.75+0.5/1.414,0.5/1.414) arc (160:23.5:1.2);
	\draw[-,dashed,very thick,color=red!30!yellow] (-5/1.06,5/2.92) arc (30:16.75:6.5);
	\draw[-,dashed,very thick,color=red!30!yellow] (2.25+0.5/1.414,0.5/1.414) arc (160:87.5:2.125);

	\draw[-,color=red] (5,0) arc (0:180:5);
	\draw[-,color=red] (1.25,0) arc (0:180:1.25);
	
	\draw[-,color=red!30!yellow] (1.75,0) arc (180:0:0.5);
	\draw[-,color=red!30!yellow] (-3.25,0) arc (0:180:0.5);
	
	\node[color=red,rotate=-90] at (0,1.25) {$\blacktriangle$};
	\node[color=red,rotate=-90] at (0,5) {$\blacktriangle$};
	
	\node[color=red!30!yellow,rotate=-90] at (2.25,0.5) {$\blacktriangle$};
	\node[color=red!30!yellow,rotate=90] at (-3.75,0.5) {$\blacktriangle$};
	
	\draw (0,0) to (0,-0.2);
	\draw (2.25,0) to (2.25,-0.2);
	\draw (-3.75,0) to (-3.75,-0.2);
	
	\node at (0,-0.4) {$0$};
	\node at (2.25,-0.4) {$c_1$};
	\node at (-3.75,-0.4) {$c_2$};
	
	\draw[->] (0,0) to (1.25/1.414,1.25/1.414);
	\node at (1.25/1.414-0.3,1.25/1.414+0.5) {$R_0$};
	
	\draw[->] (0,0) to (-5/1.414,5/1.414);
	\node at (-5/1.414+0.7,5/1.414-0.2) {$\lambda R_0$};
	
	\draw[->] (2.25,0) to (2.25-0.5/1.414,0.5/1.414);
	\node at (2.25,0.5/1.414+0.5) {$R$};
	
	\draw[->] (-3.75,0) to (-3.75+0.5/1.414,0.5/1.414);
	\node at (-3.75,0.5/1.414+0.5) {$R$};
	
	\node at (0.3,3.125) {$L_1$};
	
	\node at (1.25/2.12+1.125-0.5/2.828,0.9) {$L_2^{(1)}$};
	\node at (-1.25/2.12-1.875+0.5/2.828,1.5) {$L_2^{(2)}$};
	\node at (-4.125,1.5) {$L_2^{(3)}$};
	\node at (4,2) {$L_2^{(4)}$};
	
	\end{tikzpicture}
	\]
	\caption{The fundamental domain of the $(1,1)$ Riemann surface. The color-coded dashed lines $L_{1,2}$ are non-intersecting periodic geodesics. In this system, two of the physical parameters are lengths of the minimal periodic geodesic, while the third physical parameter is a \textit{twist}.}
	\label{figs:11pic}
\end{figure}
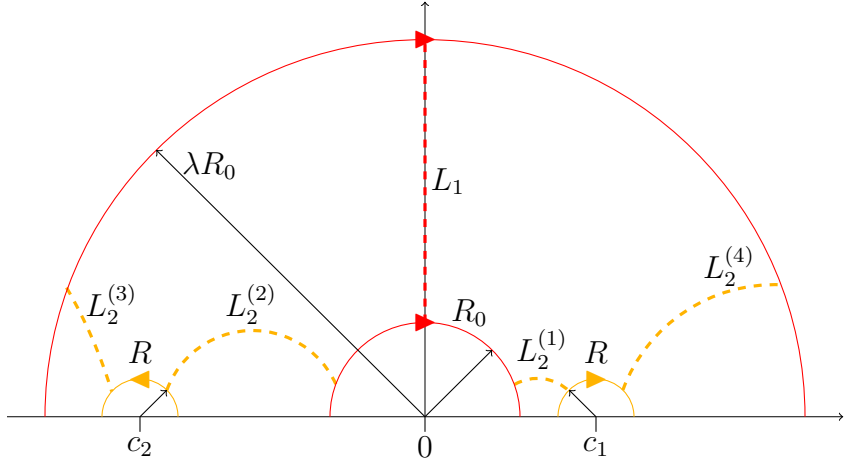

Just like the three-boundary case, the $(1,1)$-wormhole requires a two-step identification, in which the second quotient is by an orientation-reversing isometry. We show this on the $t = 0$ slice pictorially in Figure \ref{figs:11pic}, which is a more explicit version of the second picture in Figure \ref{figs:pinchfold}.

The argument is analogous to that of the three-boundary case, so we assert that, to obtain a generic $(1,1)$-wormhole with three geometrical parameters, one must first quotient by dilatation, then by the Killing vector in (\ref{KV3B}). The only difference is that the semicircle $y = \sqrt{R^2 - (x-c_2)^2}$ is on the opposite side of the fundamental domain, which means that,
\begin{align}
c_2 + R &< -R_0 \ ,\\
c_2 - R &> -\lambda R_0 \ .
\end{align}

If we introduce additional symmetry by imposing $c_1 = -c_2$, then the Killing vector in (\ref{KV3B}) reduces to a scalar multiple of $\tilde{J}_T - \tilde{J}_S = 2J_{13}$. This appears to restrict the resulting class of solutions to having only two independent parameters. In particular, if we combine $c_1 = -c_2$ with (\ref{dRConst}), then the only independent identification parameters shown in Figure \ref{figs:11pic} which appear in the Killing vectors can be taken to be $\lambda$ and $R$.

In \cite{Aminneborg:1997pz}, they take this a step further by describing a \textit{one-parameter} family of $(1,1)$-wormholes, which are obtained by taking a quotient by,
\begin{align}
\xi_1 = \alpha J_{12}\ , \quad	\xi_2 = \alpha J_{13} \ .
\end{align}
In other words, $\xi_1$ generates a particular dilatation while $\xi_2$ is a specific case of (\ref{KV3B}), for which $s = 0$. The parameter $\alpha$ is, in our language, a function of $R$. However, $\xi_1$ generates dilatation by $\lambda$ such that,
\begin{equation}
\lambda = e^{\alpha}\ .
\end{equation}

Thus, the orientation-reversing isometry for which we have found a Killing vector can be used to construct not only the three-boundary wormhole, but also the $(1,1)$-wormhole. However, note that computing the geometrical parameters in this case is less obvious.\footnote{The minimal periodic geodesic obtained from dilatation is easy, as always. However, the other minimal periodic geodesic length consists of a sum of \textit{four} terms, and the role of the twist parameter in the fundamental domain is unclear.} Regardless, there are still three independent parameters available, so our procedure should capture the full moduli space of static $(1,1)$ solutions.

%
%%%%%%%%%%%%%%%%%%%%%%%%%
\subsection{Adding More Boundaries and Genus}
%%%%%%%%%%%%%%%%%%%%%%%%%

Orientation-reversing isometries are powerful tools because we can repeat their usage in order to construct static $(n,g)$-wormholes, with $n$ boundaries and genus $g$. In order to see how, we first discuss the construction of multiboundary wormholes without any genus, \textit{i.e.} $(n,0)$-wormholes.

The construction of static $(n,0)$-wormholes is a straightforward generalization of the three-boundary case. Looking again at the three-boundary case as a two-step process, recall that quotienting by an orientation-reversing isometry was equivalent to \textit{pinching} one of the boundaries of the cylinder into two boundaries on the $t = 0$ slice. We show this pictorially in Figure \ref{figs:pinchfold}.

Each pinching corresponds to a pair of identified semicircles on the same boundary after quotienting by dilatation. We can also see that pinching always results in one additional boundary. So, to depict the fundamental domain of an $(n,0)$-wormhole at $t = 0$, we require the usual pair of concentric semicircles identified by dilatation and $n-2$ pairs of neighboring semicircles identified by an orientation-reversing isometry.

There is a small caveat to this statement, however. When considering the geometrical parameter counting, the Killing vector (\ref{KV3B}) captured the full moduli space of three-boundary wormholes by incorporating precisely two independent parameters in the coefficients. But, from (\ref{teichmuller}), we can see that pinching anything other than a cylinder adds \textit{three} geometrical parameters.
\begin{equation}
6g - 6 + 3n \xrightarrow{\text{pinching}} 6g - 6 + 3(n+1) = (6g - 6 + 3n) + 3\ .
\end{equation}

As such, (\ref{KV3B}) will not capture the full moduli space of $(n,0)$-wormholes for $n > 3$. One possible solution to this issue could be to compute the Killing vector which generates orientation-reversing isometries between circles of \textit{different} radii. We present some work related to this rather technical problem in Appendix \ref{appA}, but we leave the details to the interested reader.

Now, we are equipped to introduce genus into our scheme, thus constructing static $(n,g)$-wormholes. To do so, first we look to the construction of the $(1,1)$ geometry. Specifically, recall that quotienting by an orientation-reversing isometry in this case was equivalent to \textit{folding} two of the boundaries together at $t = 0$, as depicted in Figure \ref{figs:pinchfold}. 

Each folding corresponds to a pair of identified semicircles on different boundaries after quotienting by dilatation. Furthermore, folding will always decrease the number of boundaries by one, but increase the genus by one. Thus, we require $g$ pairs of semicircles which are on opposite sides of the central semicircles.

Again, we note that there is still a caveat to this statement. Just as for pinching, the only case in which the Killing vector (\ref{KV3B}) provides the necessary number of geometrical parameters is when quotienting two-sided BTZs to obtain $(1,1)$-wormholes. Otherwise, using (\ref{teichmuller}), folding any geometries other than a cylinder will add \textit{three} geometrical parameters.
\begin{equation}
6g - 6 + 3n \xrightarrow{\text{folding}} 6(g+1) - 6 + 3(n-1) = (6g - 6 + 3n) + 3\ .
\end{equation}

Hence, using (\ref{KV3B}) does not allow us access to the full moduli space of all $(n,g)$-wormholes, but, just as in the case of $(n,0)$-wormholes for $n > 0$, considering more general orientation-reversing isometries should do the job.

That aside, we may construct an $(n,g)$-wormhole as follows. First, perform all of the necessary quotients in order to obtain an $(n+g,0)$-wormhole. This requires quotienting by dilatation once, then by $n+g-2$ orientation-reversing isometries which correspond to pinchings. Afterwards, quotient by an additional $g$ orientation-reversing isometries corresponding to foldings. As we lose $g$ boundaries but genus increases by $g$, the resulting space is an $(n,g)$-wormhole.

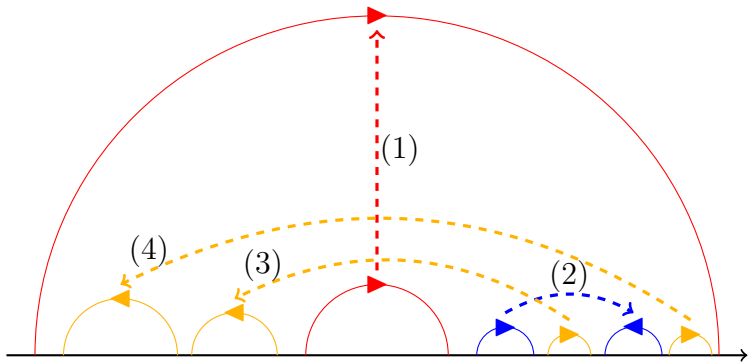
\begin{figure}
\centering
\begin{tikzpicture}[scale=0.75]
\draw[->,thick] (-6.5,0) to (6.5,0);

\draw[-,color=red] (6,0) arc (0:180:6);
\draw[-,color=red] (1.25,0) arc (0:180:1.25);

\draw[-,color=blue] (1.75,0) arc (180:0:0.5);
\draw[-,color=blue] (4,0) arc (180:0:0.5);

\draw[-,color=red!30!yellow] (3,0) arc (180:0:0.375);
\draw[-,color=red!30!yellow] (5.125,0) arc (180:0:0.375);

\draw[-,color=red!30!yellow] (-1.75,0) arc (0:180:0.75);
\draw[-,color=red!30!yellow] (-5.5,0) arc (180:0:1);

\node[color=red,rotate=-90] at (0,1.25) {$\blacktriangle$};
\node[color=red,rotate=-90] at (0,6) {$\blacktriangle$};

\node[color=blue,rotate=-90] at (2.25,0.5) {$\blacktriangle$};
\node[color=blue,rotate=90] at (4.5,0.5) {$\blacktriangle$};

\node[color=red!30!yellow,rotate = -90] at (3.375,0.375) {$\blacktriangle$};
\node[color=red!30!yellow,rotate = -90] at (5.5,0.375) {$\blacktriangle$};

\node[color=red!30!yellow,rotate = 90] at (-2.5,0.75) {$\blacktriangle$};
\node[color=red!30!yellow,rotate = 90] at (-4.5,1) {$\blacktriangle$};

\draw[->,very thick,dashed,red] (0,1.5) to (0,5.75);
\node at (0.4,7.25/2) {$(1)$};

\draw[->,very thick,dashed,blue] (2.25,0.75) to[bend left] (4.5,0.75);
\node at (3.375,1.4) {$(2)$};

\draw[->,very thick,dashed,red!30!yellow] (3.375,0.625) to[bend right] (-2.5,1);
\node at (-2,1.6) {$(3)$};

\draw[->,very thick,dashed,red!30!yellow] (5.5,0.625) to[bend right] (-4.5,1.25);
\node at (-4,1.85) {$(4)$};
\end{tikzpicture}
\caption{The fundamental domain of the $(1,2)$ Riemann surface. Step (1) is quotienting by dilatation. Step (2) is quotienting by a pinching. Steps (3) and (4) are quotienting by foldings.}
\label{figs:12pic}
\end{figure}

This simple algorithm also indicates how to draw the fundamental domain of $(n,g)$-wormholes on the $t = 0$ slice. We first consider the fundamental domain of an $(n+g,0)$-wormhole, obtained by taking one pair of semicircles identified by dilatation and $n+g-2$ pairs of semicircles identified by pinchings. Then, with the appropriate placement, we take $g$ more pairs of semicircles which will be identified by foldings. For example, Figure \ref{figs:12pic} shows the fundamental domain of a $(1,2)$-wormhole at $t = 0$.

To conclude, note that this procedure is by no means unique. For instance, to obtain, say, a $(2,1)$-wormhole, one can first construct the $(1,1)$-wormhole, then pinch the boundary afterwards. This sequence does not follow our algorithm, but still results in the same type of geometry. However, so long as each Killing vector introduces the correct number of geometrical parameters, we should capture the full moduli space of any static, $(n,g)$-wormhole with the steps described above.

Furthermore, the number of quotients is itself topologically-invariant. Define $Q(n,g)$ as the number of pinchings and foldings applied to the two-sided geometry in our algorithm. Then, by counting the steps, we have that,
\begin{equation}
Q(n,g) = (n+g-2) + g = -2 + 2g + n = -\chi \ ,
\end{equation}
where $\chi$ is the Euler characteristic of the corresponding Riemann surface. Thus, regardless of the specific steps utilized, the procedure to obtain an $(n,g)$-wormhole from the two-sided BTZ will always consist of a specific number of quotients.

%%%%%%%%%%%%%%%%%%%%%
\subsection{Introducing Rotation}\label{introRot}
%%%%%%%%%%%%%%%%%%%%%

Now, we discuss including rotation in the $(n,g)$ geometries constructed earlier. Towards that, let us begin by reviewing briefly the construction of a rotating two-sided BTZ, using a similar identification by an isometry.\footnote{The following is essentially a review of \cite{Aminneborg:1998si, Krasnov:2001va}.} The Killing vector is,
\begin{align}
\xi_{\rm rot}
& = a J_{12} + b J_{03}\nonumber\\
&= - J_1 \left( a + b \right) - \tilde{J}_1 \left( a - b \right) \nonumber\\  
&= - \left(a + b \right) \sin u \partial_u - \left( a - b \right) \sin v \partial_v \ , \label{xirot}
\end{align}
where we have closely followed the notation of \cite{Aminneborg:1998si, Krasnov:2001va} and switched to boundary light cone coordinates. We define the Killing vectors $J_1$ and $\tilde{J}_1$ in Appendix \ref{appB}.

Let us also review the strategy to read-off a non-vanishing angular velocity for the resulting geometry, since we will make explicit use of the same, later. This consists of three steps. First one finds the conformally flat boundary metric, $d\hat{s}^2$, such that $\xi_{\rm rot}$ is a Killing vector of $d\hat{s}^2$ with unit norm. This is,
\begin{equation}
d\hat{s}^2 = \frac{du dv}{\left( a^2 - b^2 \right) \sin u \sin v } \ ,\quad -\pi < u < 0 \ , \quad 0 < v < \pi \ .
\end{equation}
$\xi_{\text{rot}}$ defines a spacelike direction in the resulting quotient space, along which rotation is a symmetry. Now, we define a timelike direction in the same quotient space, by constructing a timelike Killing vector, $\xi_{\rm time}$, which is orthogonal to $\xi_{\rm rot}$. This yields,
\begin{eqnarray}
\xi_{\rm time} = - \left( a + b \right) \sin u \partial_u + \left( a - b \right) \sin v \partial_v\ ,\quad ||\xi_{\rm time}||^2 = - 1\ . \label{xitime}
\end{eqnarray}

The final step is to construct the horizon-generating Killing vector. This is done by demanding that the event horizon consists of all points which lie in the past light cone of the last point of the conformal boundary $\mathcal{I}$ (described by either $d\hat{s}^2$ or $ds^2 = - du dv$, since they belong to the same conformal class). The Killing vector is,
\begin{eqnarray}
\xi_{\rm hor} = - \sin u \partial_u + \sin v \partial_v = \frac{a}{a^2 - b^2} \left( \xi_{\rm time} - \Omega \xi_{\rm rot} \right)  \ , \label{xihor}
\end{eqnarray}
where $\Omega = b/a$ is identified with angular velocity. In fact, the choice made in (\ref{xirot}) is a special one. We will not repeat the calculation for the general case; instead, we refer the interested reader to \cite{Aminneborg:1998si, Krasnov:2001va}. The final expression of angular velocity is,
	\begin{equation}
	\Omega=\frac{\Cosh^{-1}\left(\dfrac{1}{2}\Tr\gamma^L\right)-\Cosh^{-1}\left(\dfrac{1}{2}\Tr\gamma^R\right)
	}{\Cosh^{-1}\left(\dfrac{1}{2}\Tr\gamma^L\right)+\Cosh^{-1}\left(\dfrac{1}{2}\Tr\gamma^R\right)} \ ,\label{angVel}
\end{equation} 
where $\gamma^L$ and $\gamma^R$ are exponential operators. We discuss these in more detail later.

We restrict ourselves to including rotation in the simple three-boundary and $(1,1)$ cases. We claim that the corresponding Killing vectors are of the form,
\begin{align}
\xi_1
&=a J_{12}+b J_{03}\nonumber\\
&= -(a+b)J_1-(a-b)\tilde{J}_1\nonumber\\
&= \xi_1^L + \xi_1^R \  , \label{xi1} \\
\xi_2
&=\alpha J_{12}+\beta J_{13}-b J_{02}\nonumber\\
&=-(\alpha J_1-(\beta+b)J_2)-(\alpha\tilde{J}_1+(\beta-b)\tilde{J}_2)\nonumber\\
&= \xi_2^L + \xi_2^R \ . \label{xi2}
\end{align}
Here, we have defined $\xi_i^L$ and $\xi_i^R$, $i = 1,2$, as follows.
\begin{align}
\xi_1^L &= -(a+b)J_1 \ ,\\
\xi_1^R &= -(a-b)\tilde{J}_1 \ , \\
\xi_2^L & = -[\alpha J_1-(\beta+b)J_2] \ ,\\
\xi_2^R &= -[\alpha\tilde{J}_1+(\beta-b)\tilde{J}_2] \ .
\end{align}

We proceed to justify our claim. The Killing vector $\xi_1$, upon quotienting, yields the rotating BTZ geometry, in which the angular velocity is proportional to $b/a$. Thus, we have two physical parameters: mass and angular momentum. Correspondingly, in the coefficients of just this first Killing vector, we have two free parameters: $a$ and $b$.

Now, suppose that $\alpha = 0$ in $\xi_2$. The corresponding identification yields a $(1,1)$ geometry with one rotation parameter.\footnote{To see why we obtain a $(1,1)$ geometry, look at the $t = 0$ slice, and note that the resulting Killing vector coincides with that of (\ref{KV3B}) with $s = 0$.} In the simplest case, minimal periodic geodesics intersect at angle $\pi/2$ with each other, as described in \cite{Aminneborg:1998si, Krasnov:2001va}. Furthermore, we can fix $\beta = a$. The resulting construction has two independent parameters $a$ and $b$, as does the corresponding geometry; these are the mass and the angular momentum.

On the other hand, for $b = 0$, we get back the Killing vectors that yielded the static three-boundary geometry, which is sensible since the angular velocity has been set to zero. Furthermore, in this limit, setting $\alpha = 0$ will specifically result in the static $(1,1)$ geometry discussed in Section \ref{static11}, while values of $\alpha \neq 0$ will yield a static three-boundary geometry.

Combining the observations and limits above, it is natural to think that the Killing vectors in (\ref{xi1})-(\ref{xi2}) can yield a rotating three-boundary geometry, with no genus. A naive parameter counting further supports this claim: the three-boundary rotating geometry should have three mass and one angular momentum, adding up to a total four independent parameters. The Killing vectors in (\ref{xi1})-(\ref{xi2}) clearly have four independent parameters. Setting $\beta = a$ may correspond to fixing a relation between two independent masses.

Let us now evaluate the angular velocity. We first define,
\begin{equation}
\gamma^L=[\gamma_1^L, \gamma_2^L]=e^{\xi_1^L}e^{\xi_2^L}e^{-\xi_1^L}e^{-\xi_2^L} \ ,
\end{equation}
where $\gamma_1^L$ and $\gamma_2^L$ are quantities defined in terms of the Killing vectors in (\ref{xi1}) and (\ref{xi2}),
\begin{equation}
\gamma_1^L = e^{\xi_1^L} = \exp\left(\frac{a+b}{2}\gamma_1\right) \ ,\quad \gamma_2^L = e^{\xi_2^L} = \exp\left(\frac{\alpha}{2}\gamma_1 - \frac{\beta + b}{2}\gamma_2\right)
\end{equation}
The matrices $\gamma_{1,2}$ are defined in Appendix \ref{appB}. We find,
\begin{equation}
	\frac{1}{2}\Tr\gamma^L=1-2 k^2\sinh^2\left(\frac{\sqrt{\alpha^2+(\beta+b)^2}}{2}\right)\sinh^2\left(\frac{a+b}{2}\right) \ . 
\end{equation}
where $k$ is defined by,
\begin{equation}
	k^2=\frac{(\beta+b)^2}{\alpha^2+(\beta+b)^2} \ .
\end{equation}

For $\dfrac{1}{2}\Tr\gamma^R$, we simply exchange $(a+b)\leftrightarrow(a-b)$ and $(\beta+b)\leftrightarrow(\beta-b)$. Therefore, using (\ref{angVel}),
\begin{equation}
\Omega=\frac{\sigma-\tilde{\sigma}}{\sigma+\tilde{\sigma}}
\end{equation}
where,
\begin{equation}
\sigma=\Cosh^{-1}\left[1-2 k^2\sinh^2\left(\frac{\sqrt{\alpha^2+(\beta+b)^2}}{2}\right)\sinh^2\left(\frac{a+b}{2}\right)\right] \ ,
\end{equation}
and,
\begin{equation}
\tilde{\sigma}=\Cosh^{-1}\left[1-2 \tilde{k}^2\sinh^2\left(\frac{\sqrt{\alpha^2+(\beta-b)^2}}{2}\right)\sinh^2\left(\frac{a-b}{2}\right)\right] \ .
\end{equation}
where we have defined $\tilde{k}$ by,
\begin{equation}
\tilde{k}^2 = \dfrac{(\beta-b)^2}{\alpha^2+(\beta-b)^2} \ .
\end{equation}

Hence, we obtain a single parameter $\Omega$, corresponding to equal angular velocities of the resulting geometry from the perspectives of the three different boundaries. Also note that setting $\sigma = \tilde{\sigma}$ corresponds to setting $b=0$. This will yield $\Omega = 0$, as expected from prior discussion.

As mentioned in Section \ref{riemann}, the rotation parameter does not appear at the Poincar\'e $t = 0$ slice. On this particular slice, the terms with the $b$ coefficient will vanish. Thus, from the perspective of multiboundary wormholes as Riemann surfaces being evolved through some timelike direction, we can interpret the angular momentum as arising from a lifting procedure that is alternative to that which gives us the static Killing vectors.

Furthermore, to conclude, we argue that the wormholes we have been discussing cannot have more than a single rotation parameter. This is because rotation, physically, must occur on a two-dimensional spatial slice of the overall spacetime. However, as we are considering AdS$_3$, there are \textit{only} two spatial dimensions, so all of the horizons of a multiboundary wormhole must be rotating together.

%%%%%%%%%%%
\section{Conclusions}\label{conclusion}
%%%%%%%%%%%
Different aspects of wormhole geometries as quotients of AdS$_3$ have been previously studied in the literature. However, only in the case of the two-sided BTZ was the explicit form of the Killing vector known. In this work, we have revisited the construction of three-dimensional wormholes as quotients of AdS$_3$ and, in particular, we have found the Killing vectors needed to obtain a three-boundary wormhole via the quotienting procedure. We showed that the $t = 0$ slice of the quotient space indeed captures the full moduli space of three-boundary Riemann surfaces, thus ensuring that we can construct any three-boundary static wormhole using our Killing vectors. We also present the corresponding Killing vectors for the rotating case and elaborate on how to extend our procedure to obtain both higher-boundary and higher-genus spaces. Let us point out some possible future directions related to our work:
\begin{itemize}
\item Multiboundary wormholes in spaces with less symmetry are difficult to explore. It is worth investigating if our results can be used to construct multiboundary wormholes in warped AdS$_3$. To this effect, we must understand if the Killing vector found in this work belongs to the $\text{sl}(2,\mathbb{R}) \oplus \text{u}(1)$ Lie algebra of {\em warped} AdS$_3$ Killing vectors. If this is the case, our procedure should be applicable also in warped AdS$_3$. 

\item In \cite{Fu:2018kcp}, the authors studied the holographic complexity of multiboundary wormholes. They found that, relative to an appropriate reference state, the complexity is proportional to the Euler characteristic, $\chi$. It would be interesting to understand this result in relation to the quotienting procedure. 

\item  We have outlined the procedure to generalize our result for three boundaries and zero genus to several boundaries and higher genus. It would be useful to flesh out this idea with an explicit calculation of the Killing vectors and the horizon lengths in the general case.
\end{itemize}
We plan to return to some of these questions in the near future.

%%%%%%%%%%%%%%%
\section{Acknowledgments}
%%%%%%%%%%%%%%%
We thank  Eloy Ay\'on-Beato and Aaron Zimmerman for useful discussions. EC and SS are
supported by the National Science Foundation (NSF) Grants No. PHY-1820712 and No. PHY-1620610. AKP is supported by the Council of Scientific \& Industrial Research (CSIR) Fellowship No. 09/489(0108)/2017-EMR-I. EC also thanks the Centro de Ciencias de Benasque Pedro Pascual and the organizers of the workshop ``Gravity - New perspectives from strings and higher dimensions"  for hospitality during the completion of this work.
\begin{appendices}
%%%%%%%%%%%%%%%%%%%%%%%%%%%%%%
\section{Non-Uniqueness of Identifying Semicircles}\label{appA}
%%%%%%%%%%%%%%%%%%%%%%%%%%%
	
It is noteworthy that the explicit identification of semicircles that yields a particular geometry is not unique. For example, the identification in Figure \ref{figs:3bdrypic} is not the only way to obtain a three-boundary geometry with those particular horizon lengths.

To explicitly demonstrate this, we present a different set of semicircles which, upon identification, yields a three-boundary geometry. On the $t = 0$ slice of AdS$_3$, consider the Killing vectors (taking coefficients $\lambda, a,b,c \in \mathbb{R}$),
\begin{align}
\xi_1&=\lambda \tilde{J}_D \ , \\
\xi_2&=a\tilde{J}_T + b\tilde{J}_D + c\tilde{J}_S \ ,
\end{align}

As discussed in Section \ref{KVfor3bdry}, the second Killing vector sends a point $z \in \mathbb{H}$ to $z'$,
\begin{align}
z' = \bar{b} + \frac{(z-\bar{b}) \cosh(\bar{a}\bar{c}) + \bar{c}\sinh(\bar{a}\bar{c})}{(z-\bar{b}) \sinh(\bar{a}\bar{c})/\bar{c} + \cosh(\bar{a}\bar{c})}  \ ,\label{isomC}
\end{align}
where the constants $\bar{a}$, $\bar{b}$ and $\bar{c}$ are defined by the relations (\ref{aCoeff})-(\ref{cCoeff}). Before, we eliminated the $\bar{c}$; this time, we keep it. The only constraint we impose is making $\bar{c}$ real and nonzero, so as to ensure that the Killing vector is type I$_b$.

Generically, this isometry maps geodesics to geodesics. In particular, it will map any point on a semicircle $y = \sqrt{R_1^2 - \left(x - c_1 \right)^2}$ to a point on another semicircle $y' = \sqrt{R_2^2 - \left(x' - c_2\right)^2}$. This is because, from (\ref{geodesic1}) and (\ref{geodesic2}), the only geodesics of $\mathbb{H}$ are semicircles and vertical lines, but the former do not map to the latter under (\ref{isomC}).

As such, even without imposing (\ref{constraintFinTrans}) to eliminate $\bar{c}$, quotienting the two-sided geometry by (\ref{isomC}) to pinch the space will still result in a three-boundary geometry. From our discussion in Section \ref{riemann}, even though the two Killing vectors (even on the $t = 0$ slice) have four independent coefficients, we still have three independent geometrical parameters: the lengths of the minimal periodic geodesics.

However, if we quotient by this more generic isometry beyond the three-boundary case, the presence of three independent coefficients in the Killing vector could allow us to capture the full moduli space of such static $(n,0)$-wormholes. Furthermore, (\ref{isomC}) could be written in such a way so that the related semicircles are identified by folding instead of pinching. Thus, understanding how this isometry would need to be expressed to relate arbitrary semicircles could be integral to capturing the full moduli space of any static (or, in light of the discussion in Section \ref{introRot}, rotating) $(n,g)$-wormhole. We leave pursuing this line of thought to future work.

\section{$\gamma$ Matrices} \label{appB}

In this section, we briefly review the $\gamma$ matrices that have been used in Section \ref{introRot}. The group of isometries of AdS$_3$ is SO$(2,2)$; the corresponding Lie algebra, however, is isomorphic to $\text{sl}(2,\mathbb{R}) \oplus \text{sl}(2,\mathbb{R})$. We can see this by taking linear combinations of Killing vectors that we have discussed in Section \ref{geomAdSH}, which we write in terms of both the embedding coordinates in Section \ref{geomAdSH} and the boundary light cone coordinates discussed in \cite{Aminneborg:1998si, Krasnov:2001va} and Section \ref{introRot}.
\begin{align}
&J_1=-\frac{1}{2}(J_{12} + J_{03})=\sin u\partial_u\ ,\quad && \tilde{J}_1=-\frac{1}{2}(J_{12}-J_{03})=\sin v\partial_v\ ,\\
&J_2=-\frac{1}{2}(J_{02} - J_{13})=-\cos u\partial_u\ ,\quad &&
\tilde{J}_2=-\frac{1}{2}(J_{02} + J_{13})=-\cos v\partial_v\ ,\\
&J_3=-\frac{1}{2}(J_{01} - J_{23})=\partial_u\ , \quad &&
\tilde{J}_3=-\frac{1}{2}(J_{01}+J_{23})=\partial_v\ .
\end{align}

The sets $\{J_1, J_2, J_3\}$ and $\{\tilde{J}_1, \tilde{J}_2, \tilde{J}_3\}$ form bases of copies of sl$(2,\mathbb{R})$, since the commutation relations for each of these subsets are,
\begin{align}
&[J_1,J_2] = J_3 \ , && [\tilde{J}_1,\tilde{J}_2] = \tilde{J}_3\ ,\\
&[J_1,J_3] = J_2 \ , && [\tilde{J}_1,\tilde{J}_3] = \tilde{J}_2\ ,\\
&[J_2,J_3] = -J_1 \ , && [\tilde{J}_2,\tilde{J}_3] = -\tilde{J}_1\ ,
\end{align}
Furthermore, the two copies of sl$(2,\mathbb{R})$ are distinct factors of the so$(2,2)$ Lie algebra because, for any $i,j$ from $1$ to $3$,
\begin{equation}
[J_i,\tilde{J}_j] = 0\ .
\end{equation}

Thus, we can express the $J_i$ vectors in terms of $\gamma$ matrices,
\begin{equation}
J_1=-\frac{1}{2}\gamma_1\,,\quad J_2=-\frac{1}{2}\gamma_2\,,\quad J_3=-\frac{1}{2}\gamma_0 \ , 
\end{equation}
where we define,
\begin{equation}
\gamma_0=
\begin{pmatrix}
0 && 1 \\
-1 && 0
\end{pmatrix},\quad
\gamma_1=\begin{pmatrix}
0 && 1 \\
1 && 0
\end{pmatrix},\quad
\gamma_2=\begin{pmatrix}
1 && 0 \\
0 && -1  
\end{pmatrix} \ . 
\end{equation}

\end{appendices}

\bibliographystyle{jhep}
\bibliography{multi}
\end{document}